\documentclass[fleqn,usenatbib]{mnras}

\usepackage{newtxtext,newtxmath}

\usepackage[T1]{fontenc}

\DeclareRobustCommand{\VAN}[3]{#2}
\let\VANthebibliography\thebibliography
\def\thebibliography{\DeclareRobustCommand{\VAN}[3]{##3}\VANthebibliography}

\usepackage{graphicx}	
\usepackage{amsmath}	
\usepackage{booktabs}
\usepackage{makecell}

\title[Impact of CBM on Stellar Structure \& Fate]{The Impact of Initial Mass Dependent Convective Boundary Mixing on the Structure and Fates of Massive Stars}

\author[Emily E. Whitehead et al]{
Emily E. Whitehead$^{1}$,\thanks{E-mail: e.e.whitehead@keele.ac.uk}
Raphael Hirschi$^{1,3}$,
Vishnu Varma$^{1}$,
Bernhard Mueller$^{2}$,
Federico Rizzuti$^{4}$,
\newauthor 
Cyril Georgy$^{5}$,
W. D. Arnett$^{6}$
\\
$^{1}$ Astrophysics Group, School of Chemical and Physical Sciences, Keele University, ST5 5BG, UK\\
$^{2}$
School of Physics and Astronomy, 10 College Walk, Monash University, Clayton, VIC 3800, Australia\\
$^{3}$
Kavli IPMU (WPI), University of Tokyo, 5-1-5 Kashiwanoha, Kashiwa 277-8583, Japan\\
$^{4}$Heidelberger Institut für Theoretische Studien, Schloss-Wolfsbrunnenweg 35, 69118 Heidelberg, Germany\\
$^{5}$Geneva Observatory, Geneva University, CH-1290 Sauverny, Switzerland\\
$^{6}$Steward Observatory, University of Arizona, 933 N. Cherry Avenue, Tucson AZ 85721, USA
}

\date{Accepted XXX. Received YYY; in original form ZZZ}

\pubyear{\the\year{}}

\begin{document}
\label{firstpage}
\pagerange{\pageref{firstpage}--\pageref{lastpage}}
\maketitle

\begin{abstract}
While convection has been known to play a key role in stars for many decades, its implementation in one-dimensional stellar evolution codes still represents a major uncertainty today. The purpose of this work is to investigate the impact of initial mass dependent convective boundary mixing (CBM), often referred to as overshooting, on the frequency and type of nuclear burning shell interactions that occur in low metallicity massive stars and the subsequent effect on their fates. Two grids of models were calculated using the Modules for Experiments in Stellar Astrophysics (MESA) code and a 22-isotope nuclear network, each with a different strength of CBM applied. One grid uses the typical CBM value for diffusive overshooting used in literature whereas the other grid uses CBM values guided by the results of 3D convection simulations. Interactions between the carbon, neon and oxygen shells (C-Ne-O) are common throughout both grids. The higher CBM grid also exhibits more frequent H-He and He-C interactions at lower initial masses than in the lower CBM grid. Several models also undergo multiple interaction events during evolution. While future work will be needed to fully assess the impact of the new CBM and the interactions it leads to, one expects interesting effects like unusual nucleosynthesis including more common or enhanced $i$- and $\gamma$-process nucleosynthesis. Furthermore, SN precursors and a significant change to the pre-SN structure are also expected, with many models not having the commonly expected onion-ring like structure and having a different explosion probability. 
\end{abstract}




\section{Introduction} \label{Intro}

Massive stars play a crucial role in galactic chemical evolution (GCE) by feeding material back into the interstellar medium (ISM). This happens in several different ways, the first being the strong stellar winds caused by the radiation pressure ejecting material from the stellar surface \citep{Geen2023}. Furthermore, some massive stars enrich their surroundings through violent core-collapse supernova (CCSNe) explosions that eject metals, in particular $\alpha$-elements and iron-group nuclei, into the ISM. As such it is important to understand the internal processes affecting massive stars in order to obtain a better understanding of their evolution and thus their contributions to GCE. 

One such process is convection, which is responsible for the mixing of material in different parts of the star at various stages of evolution. Despite the importance of convective and decades of studies on this topic, there are still important debates about how to best implement the inherently three-dimensional process of convection into 1D models. The most common method used in stellar evolution is the mixing length theory (hereafter MLT) proposed by \citet{Bohm_Vitense1958}, where convection is translated to one dimension by assuming that ascending and descending fluid elements travel a certain distance before they mix into the surrounding medium. The distance the fluid elements travel is known as the `mixing length' is thought to be of the order of the pressure scale height $H_{\mathrm{P}}$: $l = \alpha_{\mathrm{MLT}}H_{\mathrm{P}}$. $\alpha_{\mathrm{MLT}}$ is a free parameter close to unity and $H_{\mathrm{P}}$ is defined as the distance along the radius over which the pressure changes by a factor of $e$ and is given by \citet{MESA2011} as being equal to $P/g\rho$. However, one limitation of the MLT is that the free parameter $\alpha_{\mathrm{MLT}}$ must be calibrated empirically. Although it is often calibrated against the Sun, calibrating this variable against other types of stars, as in \citet{Joyce2018} and \citet{Valle2019}, will result in different values for $\alpha_{\mathrm{MLT}}$.  for which there does not appear to be any convergence or relation \citet{MLTreview}. 

As a fluid element approaches the convective boundary (determined in stellar evolution codes using the Ledoux or Schwarzschild criteria), its buoyancy force changes sign and the fluid element therefore decelerates \citet{Arnett2009}, `overshooting' the boundary into the stable region beyond. This causes the mixing of material across convective boundaries, which we will refer to as convective boundary mixing (hereafter CBM) in this study. This concept was studied in \citet{Saslaw&Schwarz1965}, who used a random walk approach in the first paper investigating CBM in convective cores. CBM, however, is not included in the MLT, the theory was already criticised by \citet{Renzini1987} as being unphysical \citep[see also further discussion in][]{Arnett2019}. 

Currently, CBM is commonly modeled using one of two prescriptions: step overshoot \citep{Roxburgh1965, Roxburg1978,Zahn1991} and an exponentially diffusive (exp-D) mechanism \citep{Freytag1996,Herwig2000}. 

In the step overshoot prescription, commonly used in the GENEC code for core hydrogen and helium burning phases \citep{Scott2021}, 
the CBM/overshoot region is treated as an extension of the convection zone with complete mixing of both entropy (adiabatic temperature gradient) and composition
\citep{Anders2023}. The use of this mechanism is supported by the analysis completed by \citet{Andrassy2024} for main sequence (MS) core convection \citep[see also][]{Mao2024}. An asteroseismological study by \citep{Pederson2021} also found the step overshoot prescription for CBM the best to model a majority (17 out of 26) of a group of slowly-pulsating B-type (SPB) stars. Asteroseismological measurements such as these are useful as they can provide an insight into the internal stellar structure, including the internal mixing profiles, and therefore help to constrain CBM. However, asteroseismological studies of the core structure are limited in that the late burning phases are still uncertain and it is not guaranteed that CBM is the same during the early and late phases of stellar evolution \citep{Viallet2015}. For example, it is not known if modes will propagate to the surface so late in evolution. Furthermore, the rarity of massive stars in the Universe, and thus the lack of observational data, hinders asteroseismological measurements \citep{Bowman2023}. 
On the other hand, \citet{Freytag1996} and \citet{Herwig2000} found that CBM can be effectively modelled using a diffusive approach, where the velocity of a fluid element decreases exponentially past a location close to the convective boundary (see Sect.\,\ref{CBM}). The 3D hydrodynamic models of advanced stages of shell convection in massive stars presented in \citet{Cristini2017} and \citet{Rizzuti2022} support the use of the diffusive approach, although \citet{Rizzuti2022} highlights that entropy must also be mixed in addition to the chemical composition already mixed in the diffusive mechanism. 3D models are useful to investigate the behaviour of convective motions past convective boundaries, although they are limited by computational expense to very short timescales and so are used to study specific aspects of evolution. These results are then used to develop 1D CBM prescriptions \citep[the 321D approach][]{Arnett2015} to allow models to be calculated on an evolutionary timescale to determine the effects of the new CBM prescription on the structure and evolution of the star, as was done in \citet{Scott2021} and \citet{Anders2022}.

In some cases, the mixing across the convective boundaries is extensive enough to cause interactions between nuclear burning shells. Nuclear burning shell interactions in 1D models have been studied for more than two decades \citep{Rauscher2002}, along with the changes they cause to nucleosynthetic yields \citep{Tur2009}. Yet the impact of higher levels of CBM (supported by recent 3D hydrodynamic simulations) on the occurrence of these interactions and also on their subsequent influence on further stellar evolution and ultimate fates, is still uncertain. Interactions between different nuclear burning shells have been identified. For example, \citet{Clarkson2021} find interactions between the hydrogen and helium shells in massive population III stars, whilst \citet{Roberti2025} find merging events between carbon and oxygen shells in massive stars. It is important to understand the impact that shell interactions have on a star's evolution and therefore understanding CBM is crucial for investigating the consequences for stellar structure over its lifetime.

Furthermore, studies like \citet{Temaj2024} and \citet{Laplace2025} have shown that  the strength of the CBM applied in a stellar model can affect the final fate of that star. Whereas \citet{Temaj2024} and \citet{Laplace2025} use the step overshoot implementation of CBM only applied at the boundaries of the convective hydrogen and helium cores, in this paper we investigate how changing the strength of exponential diffusive CBM applied at every convective boundary during the evolution of massive stars affects their structure, evolution and final fates.

The paper is structured as follows; Sections \ref{Methodology} outlines the grids of models calculated and discusses the 1D stellar evolution code and input physics used in this work, including the diffusive CBM mechanism used. The following Sections describe the impact that CBM has on the structure, evolution and fates of massive stars, with Section \ref{Surface&Central} discussing the surface and central properties of the models in the grids and Section \ref{InteractionSection} examining the effect on the type and incidence of nuclear burning shell interactions. The predicted fates of the stellar models in the two grids are discussed in Section \ref{FinalFates} and finally, in Section \ref{Conclusions} we present our conclusions. 

\section{Methodology} \label{Methodology}

\subsection{Convection and Convective Boundary Mixing} \label{CBM}

In this work, the Schwarzschild criterion \citep{Schwarz1958} is used to determine the location of convective boundaries and the exponential diffusive convective boundary mixing (exp-D CBM) mechanism, in which the velocity of the fluid element decays exponentially once it has passed the Schwarzschild boundary \citep{Freytag1996}, is used to model mixing past convective boundaries. This process is diffusive and is governed by the diffusion coefficient $D_{\mathrm{ov}}$ \citep{Herwig2000}, defined as

\begin{equation} \label{diffusion_coeff}
D_{\mathrm{ov}}(r) = D_{\mathrm{0}}(z_{\mathrm{0}})\mathrm{exp}\left(\frac{-2(r-z_{\mathrm{0}})}{H_{\mathrm{v}}}\right)
\end{equation}

where $r$ is the radial coordinate, $z_{\mathrm{0}} = r_{\mathrm{CB}}-f_{\mathrm{0}}H_{\mathrm{p}}$ and $r_{\mathrm{CB}}$ is the radial coordinate of the Schwarzschild boundary
(\citealt{Freytag1996} and \citealt{Kaiser2020}). $z_{\mathrm{0}}$ represents the location inside the convective zone after which the velocity of the fluid element begins to decrease exponentially and this is set by the free parameter $f_{\mathrm{0}}$. The velocity scale height, $H_{\mathrm{v}} = fH_{\mathrm{p}}$, and the free parameter $f$ represents the efficiency of this extra mixing past the convective boundary. $D_{\mathrm{0}}(z_{\mathrm{0}})$ denotes the diffusive coefficient at the location $z_{\mathrm{0}}$. In this work, we set $f_{0} = f$ and change the strength of the CBM applied by changing the value of the free parameter $f$. 
We calculate two grids of models, each with different values of the $f$ parameter applied. In the first grid (hereinafter the ``f0p02'' grid), $f = 0.02$ above the convective boundary. In the second grid, the values of $f$ are initial mass dependent, based on a factor of $\frac{1}{10}$ those calculated in Equation 8 of \citet{Scott2021} and lie in the range $0.034 \leqslant f \leqslant 0.05$. The factor of $\frac{1}{10}$ arises when converting between the step overshoot mechanism used by \citet{Scott2021} and the exp-D mechanism used in this work. This initial mass dependent set of $f$ values are referred to as `Scott-Hirschi 2021 CBM' (hereafter SH21 CBM, or simply SH21). Note that these $f$ values are specific to above (the upper edge of) the convective boundary. Values for $f$ below (the lower edge of) the boundary are a factor of $\frac{1}{5}$ of those used above the boundary in that specific model (i.e. $f_{\mathrm{below}} = \frac{1}{5}f_{\mathrm{above}}$. This is based on the 3D simulations completed by \citep{Cristini2017,Cristini2019,Rizzuti2023} finding entrainment rates approximately 5 times slower due to lower boundaries being about 5 times stiffer.

Tables \ref{tab:earlyA_table} and \ref{tab:earlyA2_table} list the $f$ values used for each model in each grid for all the upper convective boundaries (same value used for all burning phases). As explained above the $f$ values for all the lower boundaries are 1/5 the values quoted in these tables.

\subsection{MESA} \label{MESA}

The set of models in this work were calculated using revision 10398 of the Modules for Experiments in Stellar Astrophysics (MESA) stellar evolution code \citep[see][for details]{MESA2011, MESA2013, MESA2015, MESA2018, MESA2019}. This version of the MESA code is not the most recent, however it has the HDF5 output (for use in nucleosynthetic post-processing) already added. All models are non-rotating with an initial metallicity of $Z = 0.001$ and an initial mass range from 10M${_\odot}$ to 45M${_\odot}$. This wide range of initial masses is used to cover a variety of predicted fates. Each initial mass is calculated twice; once with an $f$ value of 0.02 (f0p02 grid) and again with a higher initial-mass dependent CBM value (SH21 grid). All models are calculated until the collapse of the iron core at the pre-SN stage, defined as the time when the infall velocity reaches $300\mathrm{kms}^{-1}$.

The `Henyey' implementation of the Mixing Length Theory (MLT) is used \citet{Henyey1965} with a mixing-length alpha ($\alpha_{\mathrm{MLT}}$) of 1.67 \citep{Arnett2018}.
This work uses the `Dutch' mass-loss scheme, based on the combination of mass-loss prescriptions used in \cite{Glebbeek2009}. The mass-loss rate calculated by 
\cite{deJager1988} is used up to an effective temperature of $10^{4}$K. Above this temperature, high-temperature mass loss prescriptions are used \citep[generally][on the main sequence]{Vink2001}. \cite{deJager1988} found that the rate of mass-loss increases as $T_{\textrm{eff}}$ decreases when considering stars of the same luminosity. However, when calculating the mass flux $\dot{M}/(4\pi R^{2})$ a discontinuity was found around $T_{\textrm{eff}}=10^{4}$K and the mass-loss prescription is switched to that of \cite{Vink2001}. The Vink rate is valid for effective temperatures in the range $10^{4}$K to $5 \times 10^{4}$K and a surface hydrogen mass fraction (hereafter $X_{\textrm{surface}}$) greater than 0.4. It takes into account the sudden increases in mass-loss, known as `bi-stability jumps', caused by changes in the wind ionisation. They find that the rate of mass-loss scales with metallicity as $\dot{M} \varpropto Z^{0.69}$ and $\dot{M} \varpropto Z^{0.64}$ for O stars and B supergiants respectively. If $X_{\textrm{surface}} < 0.4$, the star is considered to have become a Wolf-Rayet (WR) star and is subject to the rates described in \cite{Nugis&Lamers2000}. These limits for $T_\textrm{eff}$ and $X_{\textrm{surface}}$ are taken from \cite{Eldridge2006}, corresponding to the Humphreys-Davidson limit of stellar luminosity \citep{deJager1988} and defining the point at which the WR phase begins. 
The radiative opacities in the low temperature regime are based on the tables by \citet{Ferguson2005}, adapted to the composition used in \citet{Asplund2009}, which is used generally.

A 22 isotope network, `approx21\_plus\_Co56.net', is used to model energy generation and the nucleosynthesis of key isotopes. The newtork includes $^{1}\mathrm{H}$, $^{3}\mathrm{He}$, $^{4}\mathrm{He}$, $^{12}\mathrm{C}$, $^{14}\mathrm{N}$, $^{16}\mathrm{O}$, $^{20}\mathrm{Ne}$, $^{24}\mathrm{Mg}$, $^{28}\mathrm{Si}$, $^{32}\mathrm{S}$, $^{36}\mathrm{Ar}$, $^{40}\mathrm{Ca}$, $^{44}\mathrm{Ti}$, $^{48}\mathrm{Cr}$, $^{52}\mathrm{Fe}$, $^{54}\mathrm{Fe}$, $^{56}\mathrm{Ni}$, $^{56}\mathrm{Fe}$, $^{56}\mathrm{Cr}$ and $^{56}\mathrm{Co}$. This network contains enough isotopes to simulate the nuclear reactions along the $\alpha$-chain that occur throughout evolution until core collapse. These simulations will be `post-processed' using MPPNP \citep{Bennett2012,Pignatari2016} in order to predict complete nucleosynthesis (Aishah Harun et al, in prep.). Nuclear reaction rates are taken from the December 4$^{\mathrm{th}}$ 2017 release of the `jina reaclib', see \citet{Cyburt2010Jina}. In particular, the rate for the $^{12}\mathrm{C}(\alpha, \gamma)^{16}\mathrm{O}$ is taken from \citet{Xu2013}, the triple-$\alpha$ rate from \citet{Fynbo2005} and the $^{14}\mathrm{N}(p, \gamma)^{15}\mathrm{O}$ rate is from \citet{Imbriani2005}. 

The properties of the SH21 and f0p02 grids of models are summarised in Tables \ref{tab:earlyA_table} and \ref{tab:earlyA2_table}, respectively.



\begin{table*}
\centering
\caption{Initial properties of the models (columns 1-3) in the SH21 grid as well as properties at the end of the main sequence (4-8) and end of core helium burning (9). Columns represent the model codename (1), the initial mass $M_{\mathrm{ini}}$ (2), CBM free parameter $f_{\mathrm{CBM}}$ applied above convective boundaries (3), mass of the He-core $M_{\mathrm{He}}$ (mass coordinate where the hydrogen abundance first falls below 1\% starting from the surface, 4), main sequence lifetime t$_{\mathrm{MS}}$ (5), effective temperature Log$T_{\mathrm{eff}}$ (6), radius Log$R$ (7), luminosity Log$L$ (8), CO-core mass M$_{\mathrm{CO}}$ (mass coordinate where the helium abundance first falls below 1\% starting from the surface (9) and the central $^{12}$C mass fraction ($\mathrm{X(^{12}C)}$) (10), respectively.}
\label{tab:earlyA_table}

    \begin{tabular}{llllllllll}
    \toprule
    Name & ${M_\mathrm{ini}} [\mathrm{M_\odot}]$ & $f_{\mathrm{CBM}}$ & $M_{\mathrm{He}} [\mathrm{M_\odot}]$ & $t_{\mathrm{MS}} [\mathrm{Myr}]$ & $\mathrm{Log}T_{\mathrm{eff}} [\mathrm{K}]$ & Log$R$ [$\mathrm{R_\odot}$] & Log$L$ [$\mathrm{L_\odot}$] & $M_{\mathrm{CO}}$ [M$_\odot$] & $\mathrm{X(^{12}C)}$ \\
    \midrule

    10SH21 & 10 & 0.034 & 2.384 & 25.386 & 3.606 & 2.607 & 4.593 & 2.208 & 0.350 \\
    11SH21 & 11 & 0.037 & 2.852 & 22.040 & 3.603 & 2.672 & 4.709 & 2.667 & 0.341 \\
    12SH21 & 12 & 0.040 & 3.381 & 19.516 & 3.603 & 2.714 & 4.795 & 3.163 & 0.334 \\
    13SH21 & 13 & 0.043 & 3.916 & 17.508 & 3.600 & 2.775 & 4.903 & 3.658 & 0.329 \\
    14SH21 & 14 & 0.046 & 4.516 & 15.901 & 3.596 & 2.833 & 5.003 & 4.231 & 0.322 \\
    15SH21 & 15 & 0.05 & 5.193 & 14.650 & 3.598 & 2.873 & 5.089 & 4.785 & 0.302 \\
    16SH21 & 16 & 0.05 & 5.692 & 13.419 & 3.597 & 2.903 & 5.146 & 5.249 & 0.295 \\
    17SH21 & 17 & 0.05 & 6.221 & 12.393 & 3.598 & 2.923 & 5.190 & 5.648 & 0.285 \\
    18SH21 & 18 & 0.05 & 6.735 & 11.527 & 3.598 & 2.949 & 5.245 & 6.090 & 0.278 \\
    19SH21 & 19 & 0.05 & 7.278 & 10.790 & 3.599 & 2.969 & 5.290 & 6.592 & 0.275 \\
    20SH21 & 20 & 0.05 & 7.796 & 10.156 & 3.602 & 2.981 & 5.324 & 7.024 & 0.265 \\
    21SH21 & 21 & 0.05 & 8.350 & 9.596 & 3.605 & 2.989 & 5.352 & 7.442 & 0.257 \\
    22SH21 & 22 & 0.05 & 8.898 & 9.112 & 3.608 & 3.012 & 5.409 & 8.077 & 0.255 \\
    23SH21 & 23 & 0.05 & 9.459 & 8.689 & 3.607 & 3.025 & 5.431 & 8.476 & 0.246 \\
    24SH21 & 24 & 0.05 & 10.032 & 8.302 & 3.610 & 3.039 & 5.472 & 9.056 & 0.247 \\
    25SH21 & 25 & 0.05 & 10.581 & 7.965 & 3.618 & 3.040 & 5.505 & 9.552 & 0.241 \\
    30SH21 & 30 & 0.05 & 13.480 & 6.683 & 3.621 & 3.094 & 5.625 & 11.918 & 0.228 \\
    35SH21 & 35 & 0.05 & 16.481 & 5.856 & 3.631 & 3.129 & 5.736 & 14.591 & 0.218 \\
    40SH21 & 40 & 0.05 & 19.506 & 5.274 & 3.645 & 3.150 & 5.833 & 17.199 & 0.210 \\
    45SH21 & 45 & 0.05 & 22.610 & 4.844 & 3.683 & 3.116 & 5.917 & 19.435 & 0.211 \\
    \bottomrule
\end{tabular} 

\end{table*}

\begin{table*}
\centering
\caption{Same as Table\, \ref{tab:earlyA_table} for the f0p02 grid.}
\label{tab:earlyA2_table}
    \begin{tabular}{llllllllll}
    \toprule
    Name & ${M_\mathrm{ini}} [\mathrm{M_\odot}]$ & $f_{\mathrm{CBM}}$ & $M_{\mathrm{He}} [\mathrm{M_\odot}]$ & $t_{\mathrm{MS}} [\mathrm{Myr}]$ & $\mathrm{Log}T_{\mathrm{eff}} [\mathrm{K}]$ & Log$R$ [$\mathrm{R_\odot}$] & Log$L$ [$\mathrm{L_\odot}$] & $M_{\mathrm{CO}}$ [M$_\odot$] & $\mathrm{X(^{12}C)}$ \\
    \midrule
    10f0p02 & 10 & 0.02 & 1.949 & 23.413 & 3.672 & 2.376 & 4.394 & 1.828 & 0.339 \\
    11f0p02 & 11 & 0.02 & 2.265 & 20.109 & 3.781 & 2.236 & 4.551 & 2.101 & 0.329 \\
    12f0p02 & 12 & 0.02 & 2.608 & 17.624 & 3.700 & 2.444 & 4.643 & 2.427 & 0.313 \\
    13f0p02 & 13 & 0.02 & 2.962 & 15.687 & 3.607 & 2.709 & 4.800 & 2.782 & 0.284 \\
    14f0p02 & 14 & 0.02 & 3.314 & 14.139 & 3.607 & 2.751 & 4.881 & 3.063 & 0.261 \\
    15f0p02 & 15 & 0.02 & 3.696 & 12.892 & 3.842 & 2.308 & 4.935 & 3.307 & 0.255 \\
    16f0p02 & 16 & 0.02 & 4.096 & 11.865 & 3.917 & 2.190 & 5.001 & 3.646 & 0.245 \\
    17f0p02 & 17 & 0.02 & 4.514 & 11.002 & 4.014 & 2.025 & 5.061 & 3.962 & 0.239 \\
    18f0p02 & 18 & 0.02 & 4.936 & 10.273 & 4.048 & 1.984 & 5.114 & 4.308 & 0.233 \\
    19f0p02 & 19 & 0.02 & 5.344 & 9.647 & 3.889 & 2.328 & 5.168 & 4.783 & 0.227 \\
    20f0p02 & 20 & 0.02 & 5.782 & 9.114 & 4.003 & 2.124 & 5.214 & 5.100 & 0.227 \\
    21f0p02 & 21 & 0.02 & 6.234 & 8.645 & 4.075 & 2.003 & 5.261 & 5.402 & 0.228 \\
    22f0p02 & 22 & 0.02 & 6.684 & 8.225 & 4.054 & 2.069 & 5.306 & 5.805 & 0.224 \\
    23f0p02 & 23 & 0.02 & 7.156 & 7.864 & 4.056 & 2.081 & 5.341 & 6.181 & 0.226 \\
    24f0p02 & 24 & 0.02 & 7.604 & 7.536 & 4.011 & 2.191 & 5.380 & 6.632 & 0.222 \\
    25f0p02 & 25 & 0.02 & 8.077 & 7.240 & 3.926 & 2.381 & 5.420 & 7.114 & 0.216 \\
    30f0p02 & 30 & 0.02 & 10.520 & 6.145 & 3.715 & 2.877 & 5.568 & 9.231 & 0.211 \\
    35f0p02 & 35 & 0.02 & 12.955 & 5.421 & 3.881 & 2.611 & 5.701 & 11.041 & 0.213 \\
    40f0p02 & 40 & 0.02 & 15.574 & 4.913 & 3.627 & 3.160 & 5.782 & 13.441 & 0.200 \\
    45f0p02 & 45 & 0.02 & 18.203 & 4.540 & 3.659 & 3.142 & 5.872 & 15.548 & 0.196 \\
    \bottomrule
\end{tabular}

\end{table*}

\section{Evolution of Surface and Central Properties} \label{Surface&Central}

\begin{figure*}
	\includegraphics[width=\textwidth]{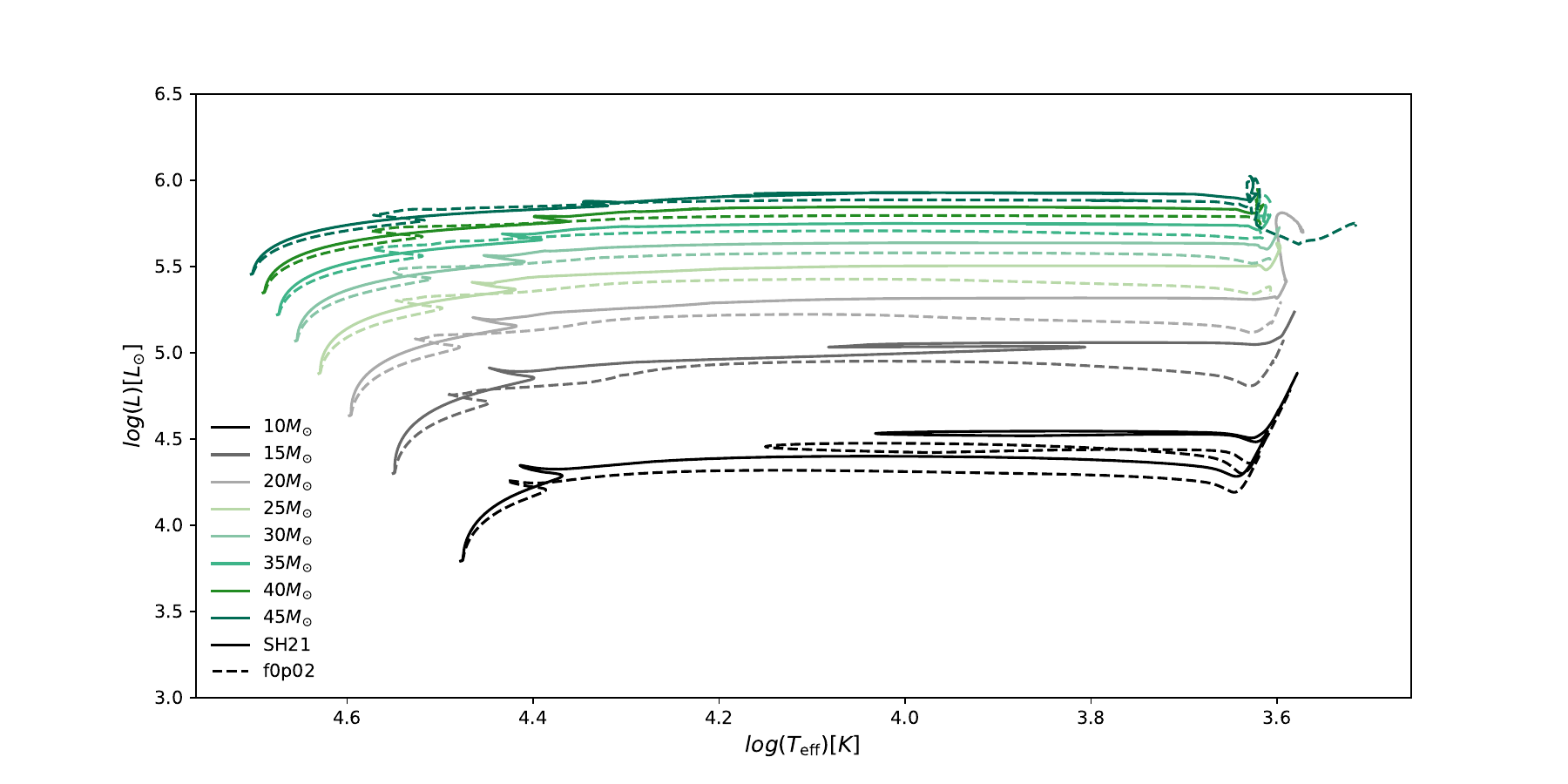}
    \caption{HRD showing the evolution of a sample of initial masses in both grids of models. Solid lines represent models in the SH21 grid whilst dashed lines show the f0p02 grid.}
    \label{fig:stackedHRD}
\end{figure*}

\begin{figure}
	\includegraphics[width=\columnwidth]{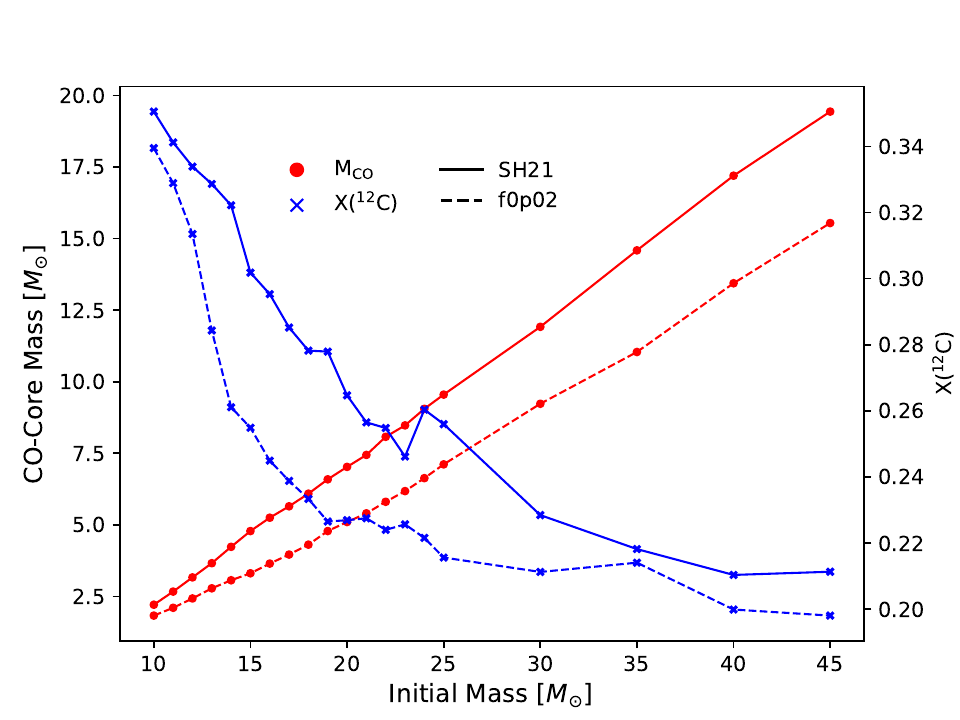}
    \caption{The mass of the CO-core ($M_{\mathrm{CO}}$) and the central $^{12}$C mass fraction (X($^{12}$C)) as a function of initial mass where $M_{\mathrm{CO}}$ and ($^{12}$C) are taken at central He-depletion. The red lines refer to the $M_{\mathrm{CO}}$ of each model whereas the blue lines represent X($^{12}$C). The  higher CBM SH21 grid is denoted by a solid line in both instances and the lower CBM strength f0p02 grid by a dashed line.
    }
    \label{fig:MCOvMini}
\end{figure}

The evolution of the surface properties of a representative sample of initial masses are presented in the Hertzsprung-Russell Diagram (HRD) in Fig.\, \ref{fig:stackedHRD}. The HRD highlights how the SH21 models behave like stars of higher initial masses when compared to their f0p02 counterparts, evolving at higher luminosities and reaching lower effective temperatures by the termination-age MS (TAMS), while at the same time having a longer MS lifetime.
This is similar to results found by \citet{Temaj2024} and \citet{Sabhahit2021} and the extended MS lifetimes are the result of the more efficient CBM in the SH21 grid of models. Indeed, this is true of all burning stages in SH21 models, where due to extra fuel being ingested into the core, the lifetime of that phase is extended. A prolonged MS leads to a higher luminosity and radius, where, because the envelope is more expanded, the effective temperature ($T_{\mathrm{eff}}$) is lower. The difference in $T_{\mathrm{eff}}$ between the SH21 and f0p02 models of a certain initial mass can be large, for example, model 25SH21 reaches the TAMS at log(T$_{\mathrm{eff}}) \approx 4.4$, whereas the corresponding model 25f0p02 reaches the TAMS at log(T$_{\mathrm{eff}}) \approx 4.5$. The difference is even larger for larger initial masses and as already found in \citet{Scott2021}, models with mass-dependent CBM better reproduce the observed MS width \citep{Castro2014}.

Furthermore, the SH21 models have consistently higher CO-core masses ($M_{\mathrm{CO}}$) at the end of core helium burning (taken at the point in evolution where the central helium abundance first drops below 1\%), as can be seen in Fig.\,\ref{fig:MCOvMini} and Tables \ref{tab:earlyA_table} \& \ref{tab:earlyA2_table}. The increase in M$_{\mathrm{CO}}$ is due to the SH21 CBM extending the core He-burning phase, meaning the mass of the resultant core is larger when the helium in the core is depleted. \citet{Kaiser2020} and \citet{Temaj2024} also find a similar relationship between the mass of the CO-core and the initial mass of the star with increasing CBM strength, lending support to the argument that enhanced mixing both above and below the convective boundaries causes the evolution of the star to take on the characteristics of those with higher initial masses up to the end of core helium burning. Both works calculate models at solar metallicity, but \citet{Kaiser2020} uses the same exponential diffusive CBM mechanism as this work whilst \citet{Temaj2024} uses the step overshoot mechanism discussed in Section \ref{Intro}, meaning that both CBM implementation have similar effect on core hydrogen and helium burning. 

Figure \ref{fig:MCOvMini} shows the M$_{\mathrm{CO}}$ increases monotonically with increasing initial mass. It also shows that the percentage increase in M$_{\mathrm{CO}}$ for each initial mass between the two grids peaks at 15M$_{\odot}$, with model 15SH21 having a CO-core mass $44.6\%$ greater than 15f0p02. Furthermore, the mass of the helium core at the end of the main sequence ($M_{\mathrm{He}}$, column 4 in Tables Tables \ref{tab:earlyA_table} \& \ref{tab:earlyA2_table}) behaves similarly explained by the larger CBM, which also leads to wider MS.

Fig \ref{fig:MCOvMini} also shows the central $^{12}$C mass fraction (X($^{12}$C)) for both grids of models, also taken at core He-depletion. As expected, X($^{12}$C) decreases as initial mass increases. This happens because, as initial mass increases, the central temperature of the star also increases and leads to the $^{12}\mathrm{C}(\alpha, \gamma)^{16}\mathrm{O}$ reaction dominating over the triple-$\alpha$ reaction. This ultimately leads to a lower X($^{12}$C) at the end of core He-burning. However, counter to intuition, the SH21 grid has a lower X($^{12}$C) for a given initial mass compared to the f0p02 grid. The values for the central $^{12}$C for both grids can be found in column 10 of Tables \ref{tab:earlyA_table} and \ref{tab:earlyA2_table}. 

\section{Nuclear Burning Shell Interactions} \label{InteractionSection}

As mentioned in the Introduction, interactions between different nuclear burning shells have long been observed in stellar evolution simulations \citep[e.\,g.][]{Rauscher2002} and several types have been identified. In this section we discuss the interactions found in the two grids computed in this work. These interactions can happen between a number of different burning shells and the specific shells involved determine the evolutionary stage and mass coordinate within the stellar interior at which the interaction occurs. As a consequence of the more efficient CBM applied at every convective boundary, the SH21 grid is expected to contain frequent nuclear burning shell interactions and this is indeed the case (See Tables \ref{tab:finalA_table} and \ref{tab:finalA2_table} which identify for all models  the types of nuclear burning shell interactions that they undergo). It is important to also note that the f0p02 grid also contains frequent interactions, although there are differences in the types of interactions that occur between the two grids. The SH21 grid also contains models that undergo `diving' shell events occurring in relatively early evolutionary stages and the significance of these events for the remainder of evolution, and likely fate, is discussed in Section \ref{DivingHe-C}. Whereas most stellar models undergo only a single interaction event during their evolution, several models see two or more instances of interacting shells and examples from the SH21 grid are discussed in Section \ref{MultiEvents}. The interactions we observe in the two grids can be separated into two approximate categories: those involving hydrogen, helium or carbon burning shells as Early Burning Phase (EBP) interactions and those involving carbon, neon, oxygen or silicon burning shells as Late Burning Phase (LBP) interactions. 
EBP interactions generally occur before the onset of convective core oxygen-burning and are discussed in Section \ref{EPI}. 
LBP generally occur after that point and are discussed in Section \ref{LPI} respectively.

\subsection{Early Burning Phase (EBP) Interactions} \label{EPI}

Throughout the SH21 grid, we see EBP interactions. These occur before the end of core oxygen-burning and they involve either the entrainment of hydrogen into the convective helium shell (H-He), or mergers between the helium and carbon shells (He-C). Some models containing an He-C interaction also undergo a diving He-shell relatively early in their evolution, although their cause is uncertain and needs more detailed investigation as discussed below.

\begin{figure*}
	\includegraphics[width=\textwidth]{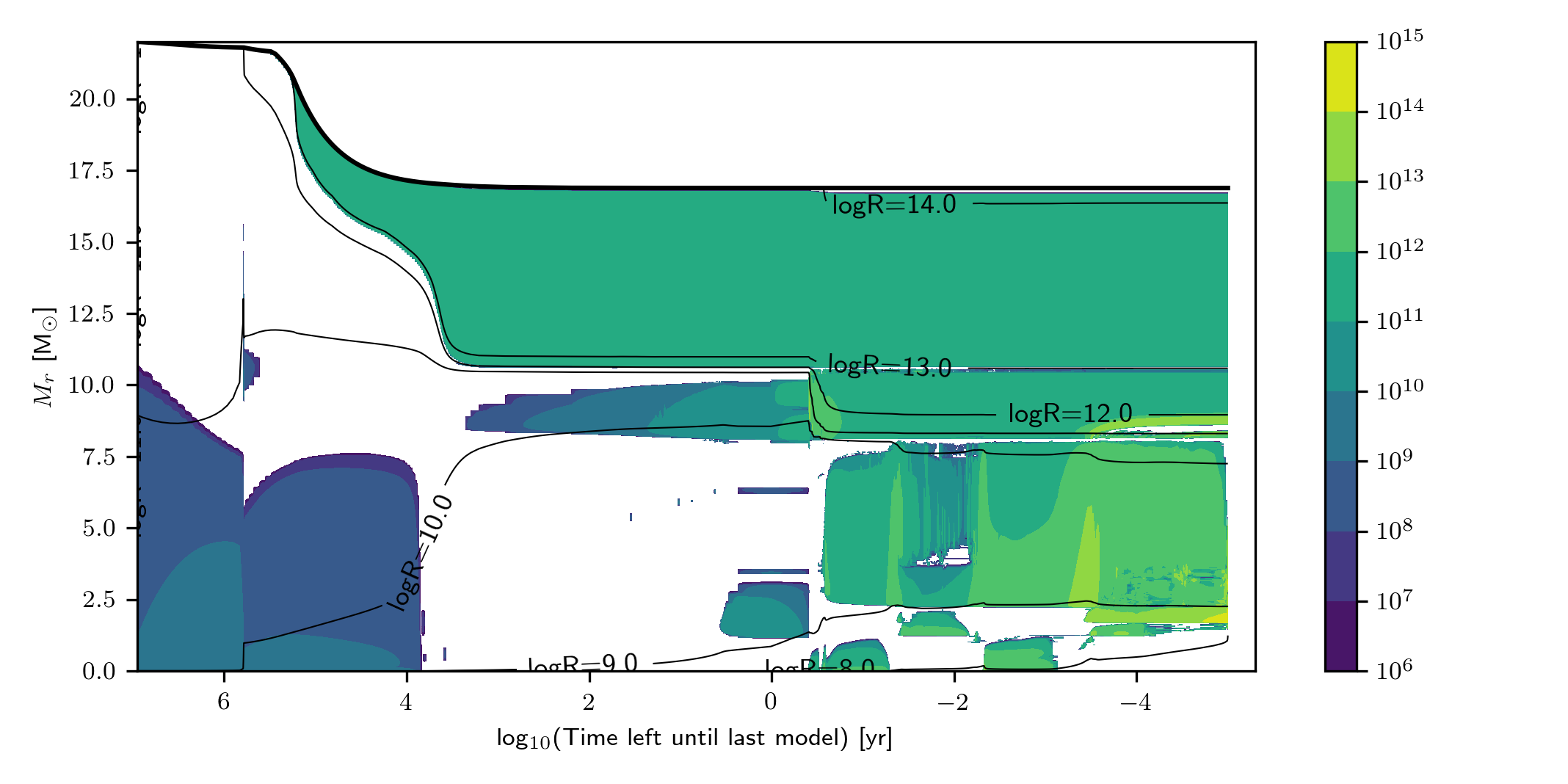}
    \caption{Structure-evolution (Kippenhahn) diagram for the 22SH21 model, showing the convective structure along the mass coordinate (M${_\odot}$) as a function of the Log of time-left before core collapse (years). The colour bar represents the specific turbulent kinetic energy in convective regions, $ v_{\mathrm{conv}}^2/2$ [cm$^2 s^{-2}$] and the isoradius contours highlight the expansion (when contours go down or contraction when contours go up in mass coordinates) of the different layers of the stellar interior over the course of evolution. The diagram shows that, just before the onset of core O-burning, expansion is triggered by the entrainment of hydrogen into the He-shell. It can also be seen that the convective velocities are higher after this event, as shown by the more yellow area in this expanded region.}
    \label{fig:2D_maps_22SH21}
\end{figure*}

\begin{figure*}
	\includegraphics[width=\textwidth]{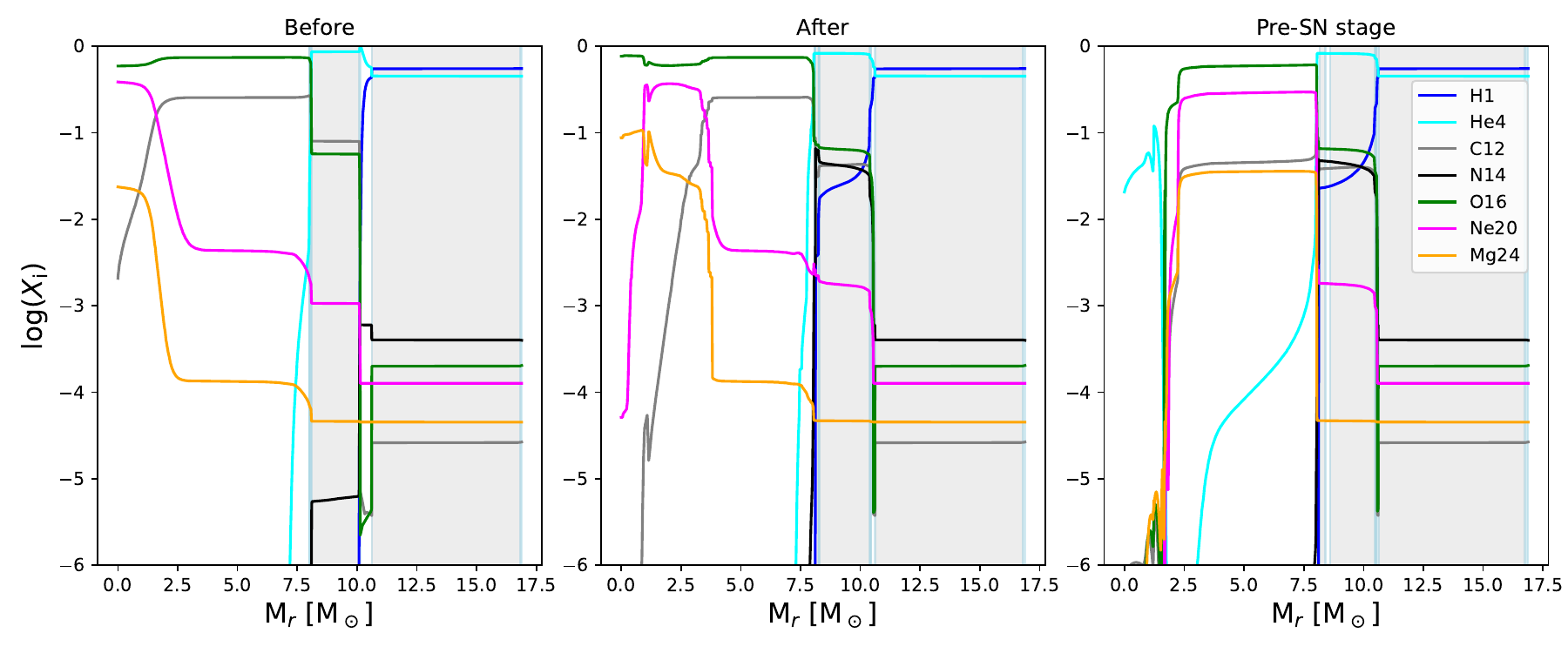}
    \caption{Abundance profiles taken before (left), during (middle) and after (right) the H-He interaction in the 22SH21 model. The grey shaded areas show convective regions and the blue shaded areas represent the CBM regions. During the interaction, hydrogen (dark blue line) is entrained into the He-shell (light blue line), being brought down to a mass coordinate of $\approx 8\mathrm{M}_\odot$.}
    \label{fig:22SH21_abundance}
\end{figure*}

\begin{figure*}
	\includegraphics[width=\textwidth]{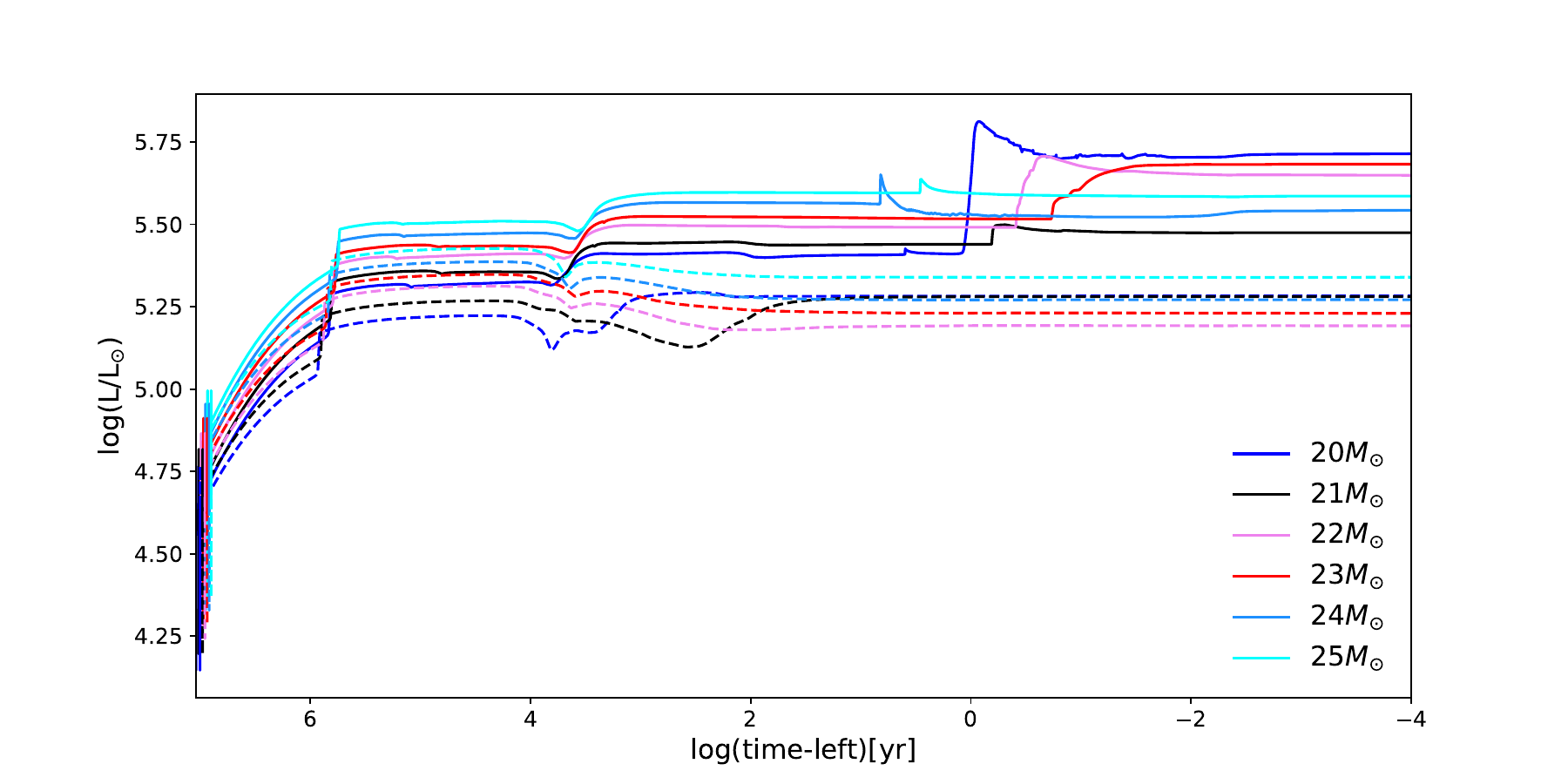}
    \caption{Profile of the Log of the luminosity against the Log of the time-left before collapse for a sample of models that undergo H-He interactions in the SH21 grid (solid lines) and their lower CBM counterparts in the f0p02 grid (dashed lines). All models undergoing H-He interactions show a marked increase in luminosity at approximately the same time as they undergo H-He interactions in their interiors, showing that the expansion and {increase in energy generation} caused by such interactions are extensive enough to affect the surface of the stars.}
    \label{fig:LvTimeleft}
\end{figure*}

\subsubsection{H-He Interactions} \label{H-He}

H-He interactions are common in the SH21 grid in models with initial masses in the range 20-25M${_\odot}$. They also occur in the 35M${_\odot}$ and 45M${_\odot}$ models in the f0p02 grid. In Fig.\,\ref{fig:2D_maps_22SH21}, we show an example of an early phase H-He interaction in the 22M${_\odot}$ model of the SH21 grid in a Kippenhahn diagram. Such diagram shows the evolution of the structure of the star along the mass coordinate as a function of the Log of the time-left until the collapse of the star. The coloured regions are convective and the black isoradius contours highlight the radial changes in mass coordinate throughout evolution, where the radius is in units of centimetres. The $\mathrm{logR=11.0}$ (in between 10.0 and 12.0) contour shows the expansion caused by the early phase interaction at a $\mathrm{Log(time-left)} \approx -0.5$ where it drops from a mass coordinate of $\approx 10.2$M${_\odot}$ to $\approx 8$M${_\odot}$.
This is a result of hydrogen being entrained into the convective helium shell below it. This entrainment also causes the kinetic energy of the convective zone to increase by 1 order of magnitude, as shown by the colour bar. This early phase interaction can be seen in the abundance profiles in Fig.\,\ref{fig:22SH21_abundance}, where the mass fractions of key isotopes are shown as a function of the mass coordinate. The three abundance profiles are taken before and after the onset of H-entrainment and also at the end of evolution (pre-SN stage).
The grey shaded regions represent the convective regions of the star at the point in evolution the abundance profile is taken at whereas the blue shading represents CBM regions. Hydrogen is represented by the dark blue line and the middle panel of Figure \ref{fig:22SH21_abundance} shows how the hydrogen is being significantly entrained into the helium shell (light blue line) in the mass coordinate range $\sim$8 - 10 M${_\odot}$. The abundance profile taken at the pre-SN stage shows that this entrainment persists, with hydrogen being brought down to a mass coordinate of $\sim 8$M${_\odot}$. Whilst our models only follow a small list of isotopes, H-He interactions appear to be a possible site for the intermediate (in between the slow and fast processes) neutron-capture process ($i$-process) nucleosynthesis. 
Suitable conditions for the $i$-process have been found in AGB stars \citep[e.\,g.][]{Cristallo2011,Choplin2021, Remple2024} as well as in massive stars \citep[e.\,g.][]{Clarkson2018,Clarkson2021}, in both cases generally at very low and zero metallicities.
The $i$ process in our models will be the topic of a future study (Aishah Harun et al, in prep) that will post-process all the models from this work and produce a full nucleosynthetic analysis of the interactions but preliminary results suggest that H-He interactions our models at $Z=0.001$ provide suitable conditions for the $i$ process, increasing the maximum metallicity at which this process can occur.
Furthermore, \citet{Pignatari2025} suggests that, if a supernova shock reaches a merged H-He shell, $^{26}$Al and $^{22}$Na may be produced as a result of the high temperatures, with implications for pre-solar grains.

Figure \ref{fig:LvTimeleft} shows the evolution of the luminosity for the 20-25M${_\odot}$ models in both grids as a function of the time left before the collapse of the core. It demonstrates the impact of the H-He interactions on the surface properties of the star. Once again taking the 22SH21 model as an example, there is a luminosity increase of $\sim77.8\%$ at approximately the same time that the H-entrainment begins. This significant increase in luminosity highlights not only the structural changes caused by the H-He interaction, but also the significance of the time and mass coordinate that it occurs at. This is because H-He interactions happen both early enough in the evolution and shallow enough within the stellar interior (within a mass coordinate range $7.5 \lesssim M/M_{\odot} \lesssim 19$) that feedback can reach the surface. These surface effects, the luminosity and corresponding radius increase, persist until the core collapse, meaning that they could possibly explain supernova precursor events \citep[see e.\,g.][]{Bruch2021}.

These H-He interactions are triggered by the core contraction in between core burning stages. This is because as a core burning stage ends, so does the related nuclear energy release. This lack of support against the gravitational force leads to the core and any nuclear burning shells resuming contraction. As a result, temperature and energy generation increase at the bottom of most shells and cause the shells to grow in mass. When an H-He interaction occurs, the energy generation increases significantly as the entrained hydrogen is burnt at the higher He-burning temperatures and therefore the layers above the interaction site expand. H-He interactions are more likely to occur in stars of low metallicity because their lower CNO abundances mean that hydrogen burning occurs at a higher temperature. This means that there is a lower entropy barrier between the hydrogen and helium layers and hydrogen entrainment therefore occurs more easily. 
He-H interactions have often been found in the literature in popIII models ($Z=0$).
\citet{Heger2010} find evidence of interactions between the hydrogen and helium shells in their zero metallicity, non-rotating 15M${_\odot}$ and 25M${_\odot}$ models. \citet{Limongi2012} and \citet{Clarkson2018, Clarkson2018erratum}. 

\citet{Ritter2018} also find hydrogen ingestion events in several models, namely the 20M${_\odot}$ at $\mathrm{Z} = 0.001$ model, although the H-He interaction occurs very late in evolution, during silicon shell burning. Other examples include 20 \& 25M${_\odot}$ models at $\mathrm{Z} = 0.0001$, where the interaction takes place much earlier, at the onset of carbon shell burning. This is more comparable to the results found in this work where, for example, the H-He interaction in the 22SH21 model begins at the end of the first carbon shell. These previous studies combined with our new models using higher CBM likely imply H-He interactions to be more frequent (possibly very frequent) in the early Universe than previously thought.

\subsubsection{Diving He-Shells and He-C Interactions} \label{DivingHe-C}

\begin{figure*}
    \includegraphics[width=\textwidth]{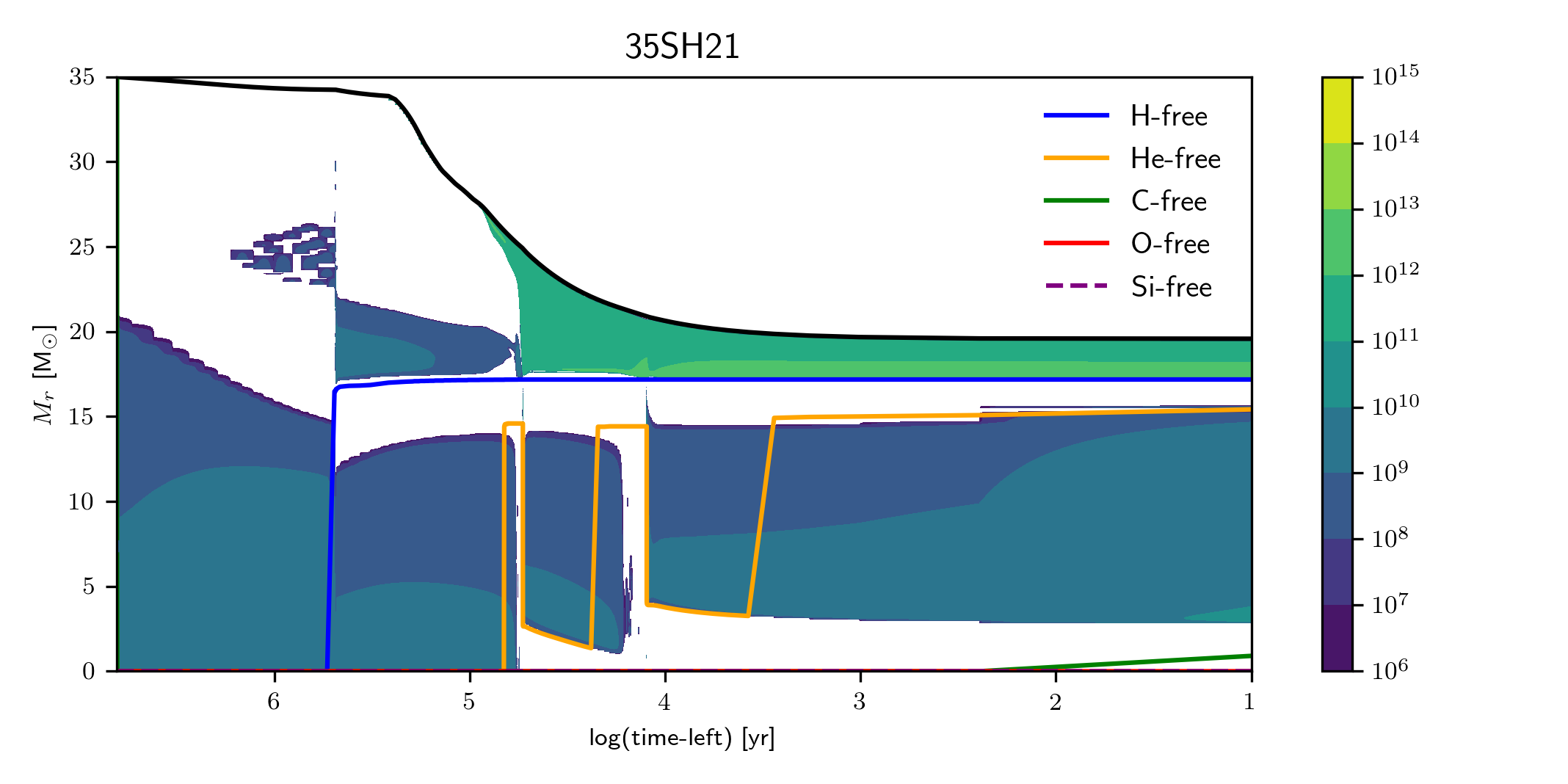}
    \caption{Kippenhahn diagram for the 35SH21 model. The `X'-free area inside each of the lines represents the regions where the abundance of isotope X is less than 1\%. Similarly to Fig 3, the colour bar represents the specific turbulent kinetic energy in convective regions, $\frac{1}{2} v_{\mathrm{conv}}^2$ [cm$^2 s^{-2}$]. In the range $4.3 \lesssim \mathrm{log(time-left)[yr] \lesssim 4.7} $ there is a diving convective He-shell shown by the He-free line in orange. At $\mathrm{log(time-left)[yr] \approx 4} $, helium is mixed down in to the convective carbon shell in a He-C interaction that persists until the pre-SN stage.}
    \label{fig:35SH21_massfrac}
\end{figure*}

\begin{figure*}
	\includegraphics[width=\textwidth]{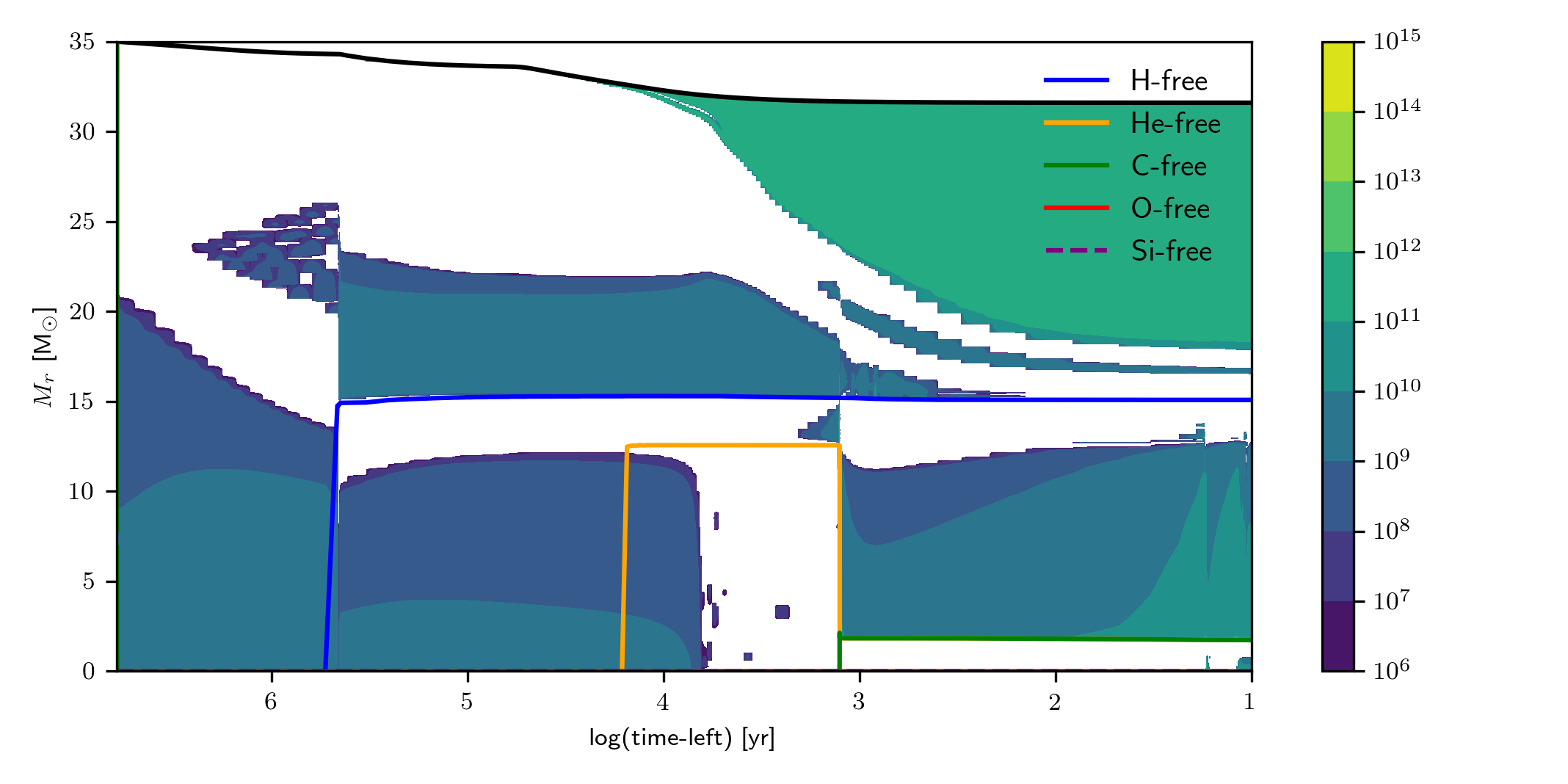}
    \caption{Kippenhahn diagram for a model with an initial mass of 35M${_\odot}$, the same metallicity as the SH21 and f0p02 grids and an $f$ parameter above convective boundaries of $f_{\mathrm{CBM,above}} = 0.035$ and below convective boundaries of $f_{\mathrm{CBM,below}} = 0.007$. In this model there is no diving He-shell, suggesting that it is suppressed when a lower strength CBM is used.}
    \label{fig:f0p035massfrac}
\end{figure*}

\begin{figure*}
    \includegraphics[width=\textwidth]{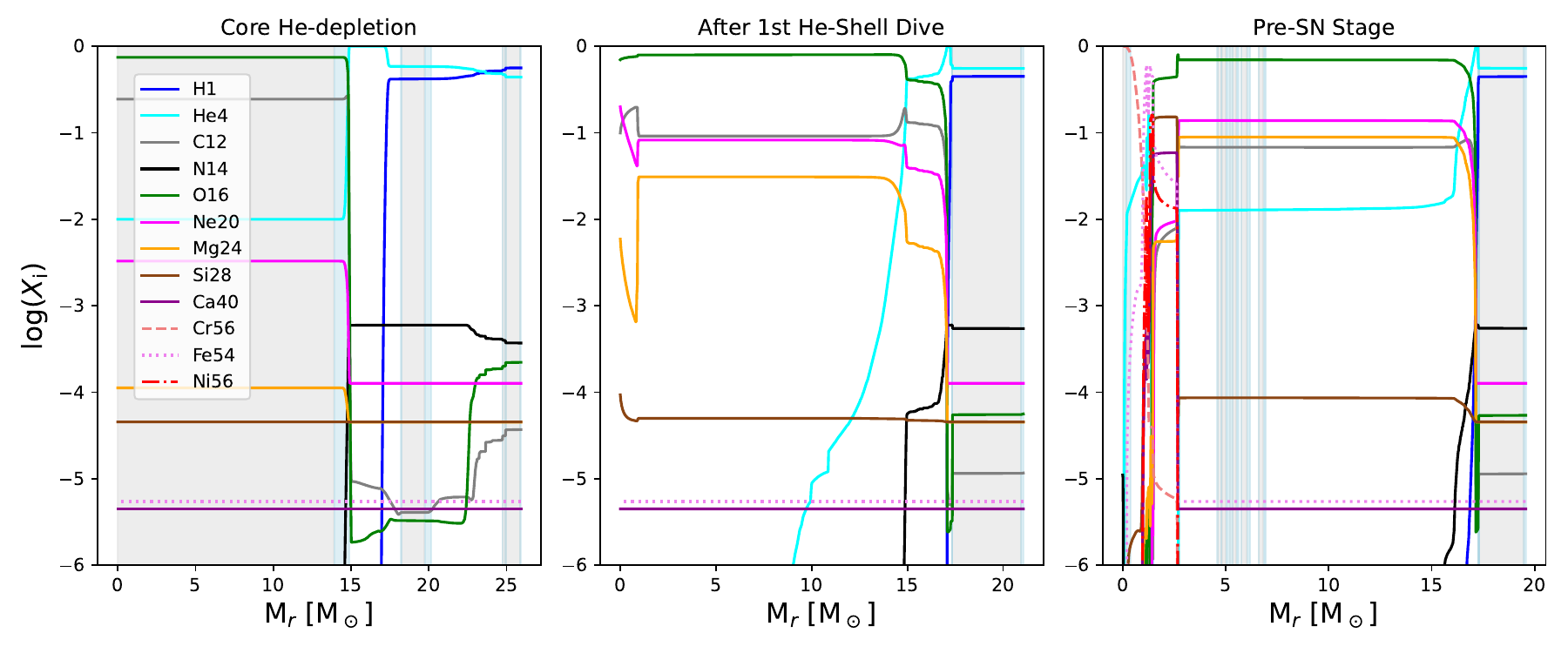}
    \caption{Abundance profiles for the 35SH21 model, showing the structure before the He-shell dive, at core He-depletion in the left panel. The middle panel shows the structure after the He-shell dives and the right-hand panel shows the pre-SN structure. }
    \label{fig:35SH21_HeC}
\end{figure*}

There is also a subset of the SH21 grid, in the initial mass range 30-45\,M${_\odot}$, that experience a dramatic diving of the He-shell early in evolution. This is shown in Fig.\,\ref{fig:35SH21_massfrac}, where the diving He-shell begins $\approx 10^{5}$ years before the collapse of the core and reaches down to 1.5\,M${_\odot}$. This phenomenon has a large impact on $M_{\mathrm{CO}}$, as can be seen in Tables \ref{tab:finalA_table} \& \ref{tab:finalA2_table}, with the 35SH21 model having an $M_{\mathrm{CO}} \approx 24.9\%$ of that of its lower mixing counterpart 35f0p02 at the pre-SN stage. Furthermore, the $M_{\mathrm{CO}}$ of the 35SH21 model taken at the pre-SN stage is only $\approx 18.6\%$ of that taken at the end of core He-burning (Table \ref{tab:earlyA_table}). The significant reduction in the CO-core mass has implications for the final fates of the stars and is discussed in Section \ref{FinalFates}.  Figure \ref{fig:35SH21_HeC} shows a series of abundance profiles taken throughout the evolution of the 35SH21 model (the same model shown in Figure \ref{fig:35SH21_massfrac}) and highlights the impact of the diving shell on the abundance profiles. The first panel shows the abundance profile taken at the very end of core He-burning, whereas the second panel shows the abundance profile taken after the first He-shell dive. Here we can see that helium has been brought down but then burnt and only left above a mass coordinate $\approx 9\mathrm{M}_{\odot}$ by the recently ended diving He-shell. The last panel in Figure 8 shows the abundance profile taken at the pre-SN stage and shows how helium has been mixed down into the carbon shell to such an extent (down to $\approx 3.5\mathrm{M}_{\odot}$) that the shell becomes a totally mixed He-C shell. This He-C interaction can be seen in Fig \ref{fig:35SH21_massfrac}, beginning $\approx 10^4$ years before collapse.

In order to test the dependence of the diving He-shell on the CBM free parameter $f_\mathrm{CBM}$, a test model using an initial mass of 35\,M$_{\odot}$ and $f_{\mathrm{CBM}}=0.035$ above convective boundaries and $f_{\mathrm{CBM}}=0.007$ below was calculated and Fig.\,\ref{fig:f0p035massfrac} shows how the small reduction in $f_\mathrm{CBM}$ inhibits the He-shell dive. It is thus unclear how realistic such diving shells are and what causes them. This will require further investigation beyond the scope of this work.
This being said, models that experience a diving He-shell also undergo an He-C interaction later on, another type of early phase interaction that involves the convective helium and carbon shells with a similar impact. The test model using $f_{\mathrm{CBM}}=0.035$ still experiences that He-C interaction (a bit more than 1000 years before collapse) and such interactions will warrant further investigations in future studies as they have not been discussed in the literature and they strongly modify the inner structure of stars above 30\,M$_\odot$.

\subsection{Late Burning Phase (LBP) Interactions} \label{LPI}

\begin{figure*}
	\includegraphics[width=\textwidth]{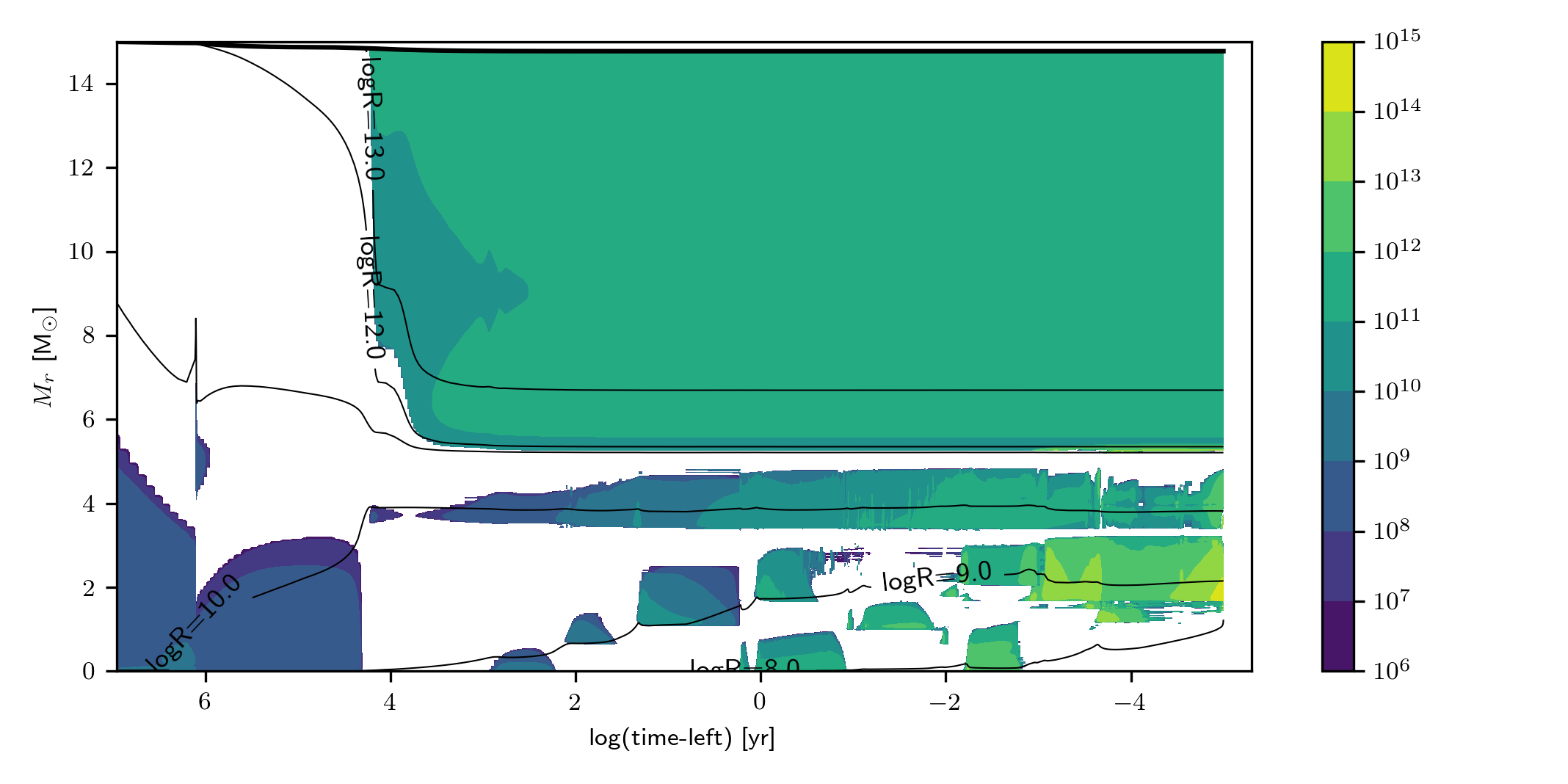}
    \caption{Kippenhahn diagram for the 15f0p02 model that shows a C-Ne shell merger at a $\mathrm{log(time-left)} \approx -2 $ and a mass coordinate of $\approx 2$M${_\odot}$.}
    \label{fig:2D_maps_15f0p02}
\end{figure*}

\begin{figure*}
	\includegraphics[width=\textwidth]{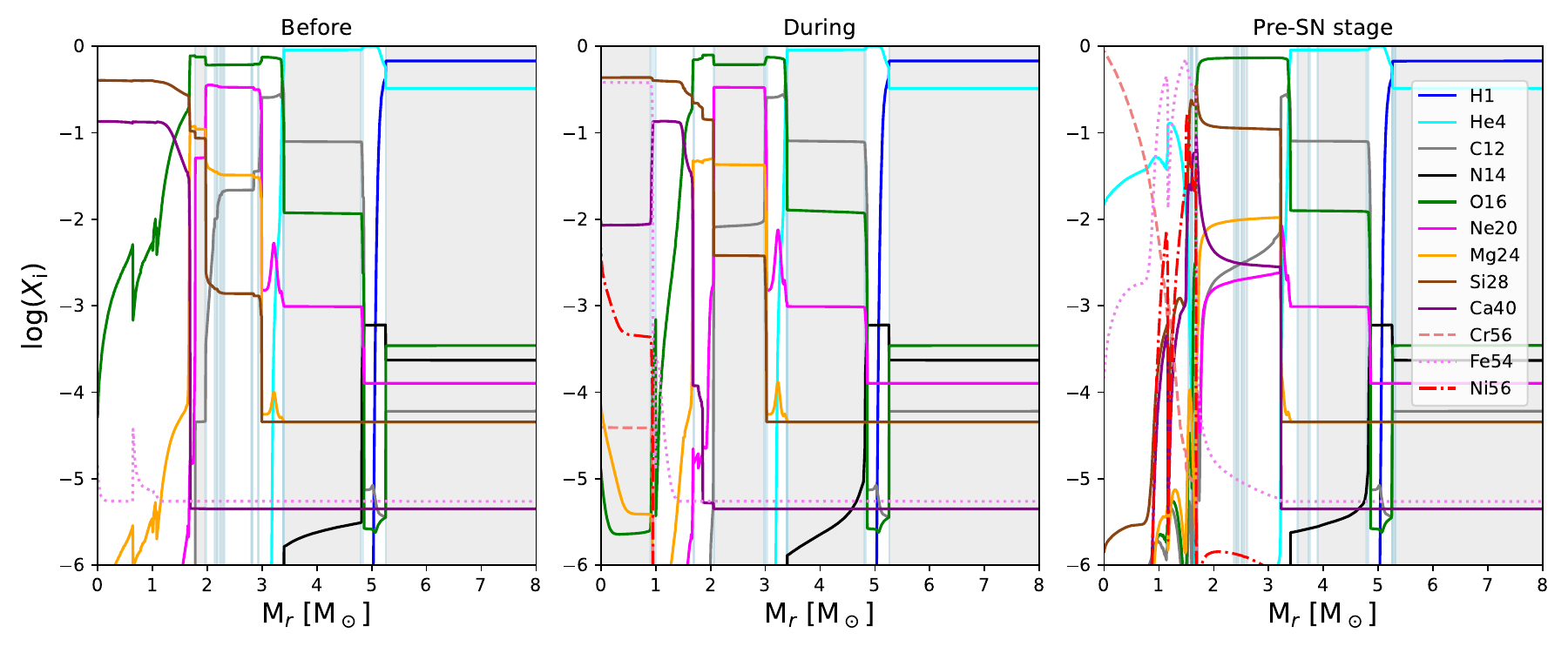}
    \caption{Abundance profiles for the 15f0p02 model that show the mass fractions before (left), during (middle) and after (right) the interaction between the carbon and neon shells. The middle panel shows how the carbon and neon burning fronts align at a mass coordinate of approximately 2M${_\odot}$. The abundance profile in the right-hand panel, taken at the pre-SN stage, shows that this LBP interaction persists until the end of evolution.}
    \label{fig:15f0p02abundance}
\end{figure*}

\begin{figure*}
	\includegraphics[width=\textwidth]{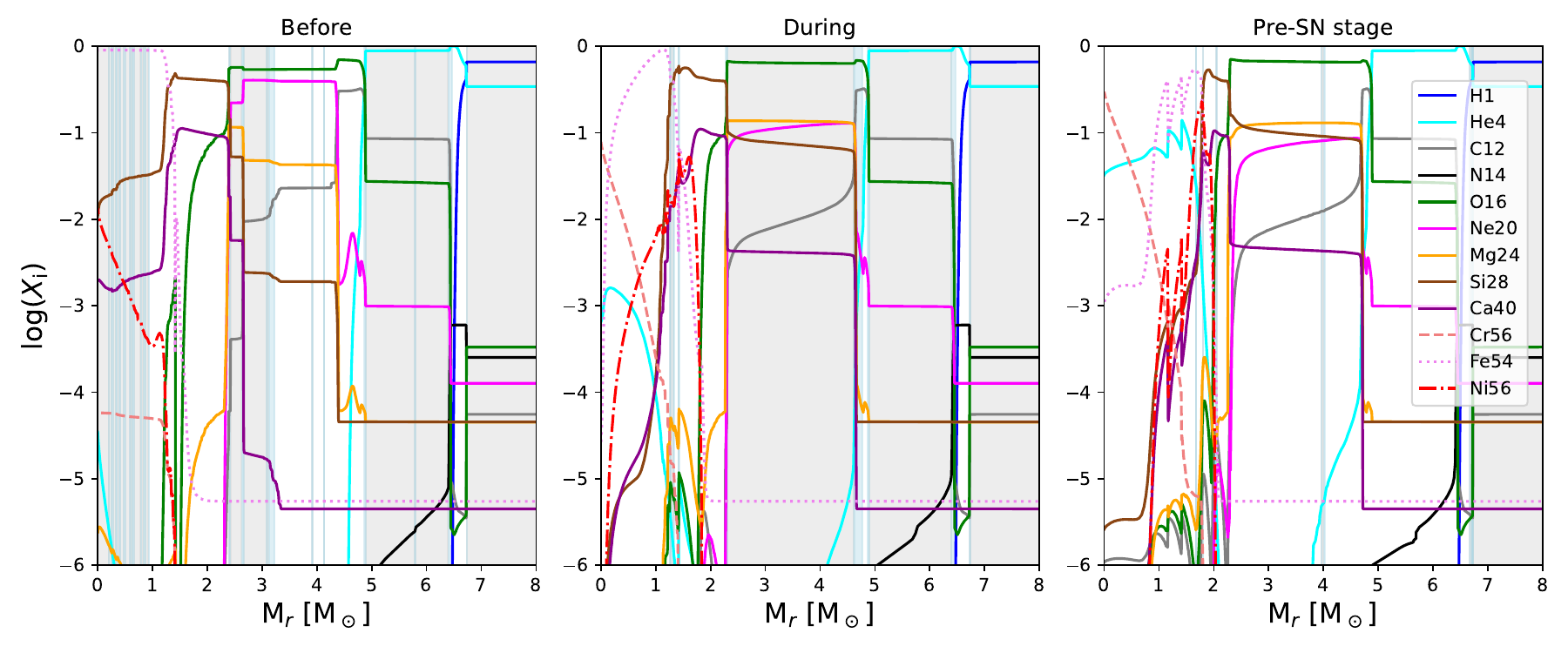}
    \caption{Abundance profiles for the 15SH21 model that illustrate the structure before (left), during (middle) and after (right) the C-Ne-O interaction. At the pre-SN stage it can be seen that the burning fronts for carbon, neon and oxygen have aligned at $\approx 2.2\mathrm{M}_\odot$. The abundance of silicon, a product of oxygen burning, has also increased in this region ($2.2 \lesssim M/M_{\odot} \lesssim 4.8$), signalling an interaction has occurred.}
    \label{fig:15SH21abundance}
\end{figure*}

LBP interactions between several different nuclear burning shells are found in a large proportion of models in both grids after the onset of core oxygen burning. They mainly comprise interactions between the C- and Ne-shells or the C-, Ne- and O-shells, although there are also single cases of Si-O and Ne-O interactions. LBP interactions happen deep within the stellar interior and, unlike their EBP counterparts, late in evolution. This means that the surface does not have time to react to the interaction event and therefore no surface effects such as radial or luminosity increases are seen. In the SH21 grid, LBP interactions occur in the initial mass range $10 \leq M/\mathrm{M}_{\odot} \leq 19$. On the other hand, LBP interactions are seen over the whole initial mass range in the lower mixing f0p02 grid, with C-Ne mergers being more frequent. The evolution of the 15f0p02 model is shown in Fig.\,\ref{fig:2D_maps_15f0p02} where the Kippenhahn diagram shows that an interaction between the carbon and neon shells occurs $\approx 10^{-2}$ years before core collapse around a mass coordinate of 2\,M${_\odot}$. This interaction can also be seen in the abundance profiles in Fig.\,\ref{fig:15f0p02abundance} where, moving from the first to second panel, the carbon and neon burning fronts align at a mass coordinate of 2\,M${_\odot}$. At this point in evolution, the two shells have merged. The abundance profile taken at the end of evolution (right-hand panel) shows that the signature of the merger is still present at the pre-SN stage. C-Ne mergers are also present in \citet{Heger2010} in two models from their zero metallicity grid, seen in the pre-SN abundance profiles in figure 4 of that paper. In comparison, our f0p02 model's higher mixing counterpart, 15SH21, contains a LBP interaction where the carbon and neon nuclear burning shells also merge with the oxygen shell as shown in Fig.\,\ref{fig:15SH21abundance}. The C-Ne-O interaction occurs at $\approx 2.2$M${_\odot}$ (middle panel) and persists until the end of evolution (right-hand panel).
While preliminary post-processing calculations with a full network do not indicate that C-Ne-O interactions lead to $i$-process nucleosynthesis, such interactions have been observed before in different studies \citep[e.\,g.][]{Ritter2018}. In particular, recent work by \citet{Roberti2025} shows that such mergers could explain abundances for elements like potassium and scandium as well as initiate the $\gamma$-process before the star explodes \citep{Roberti2025l}.

\subsection{Multiple Interaction Events} \label{MultiEvents}

\begin{figure*}
	\includegraphics[width=\textwidth]{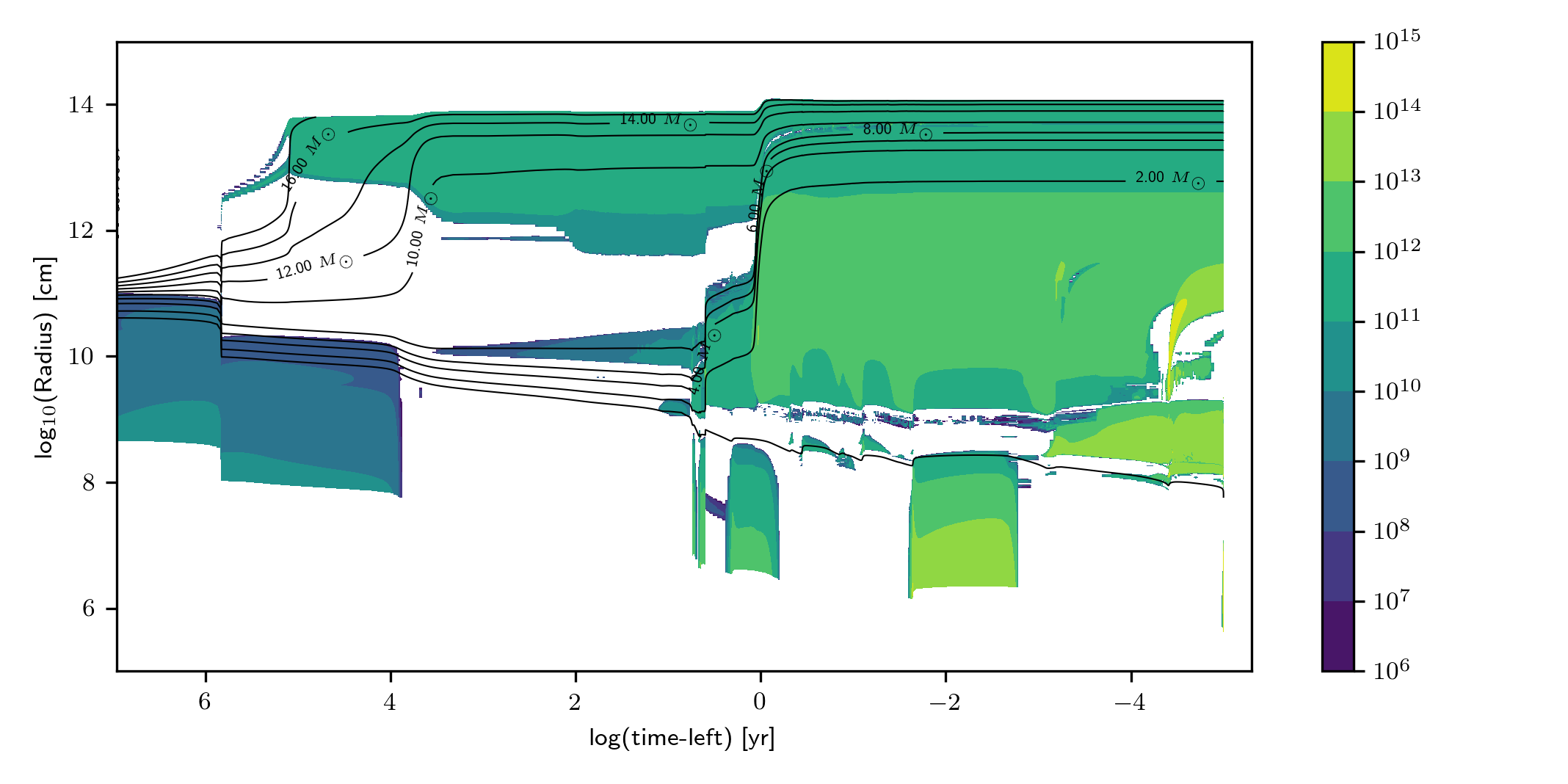}
    \caption{Kippenhahn diagram of the 20SH21 model showing the convective structure using the Log of the radius in centimetres as a function of the Log of the time-left before core collapse in years. The colour bar represents the turbulent kinetic energy in convective regions, $\frac{1}{2} v_{\mathrm{conv}}^2$ [cm$^2$] and the isomass contours highlight the expansion (when isomass contours go up in radius) of the different layers of the stellar interior over the course of evolution. The 2\,M${_\odot}$ isomass contour shows the expansion caused by an He-C interaction that occurs at a $\mathrm{log(time-left)} \approx 0.5$. The second instance of expansion, also at a mass coordinate of 2\,M${_\odot}$ is caused by the entrainment of hydrogen into the helium shell. This H-He interaction results in the 2\,M$_{\odot}$ isomass contour increasing from a $\mathrm{Log(Radius)} \approx 10$ to $\mathrm{log(Radius)} \approx 12.5$, with the expansion propagating all the way up to the surface and showing an increase in convective velocities of approximately one order of magnitude.}
    \label{fig:2Dmaps20SH21}
\end{figure*}

\begin{figure*}
	\includegraphics[width=\textwidth]{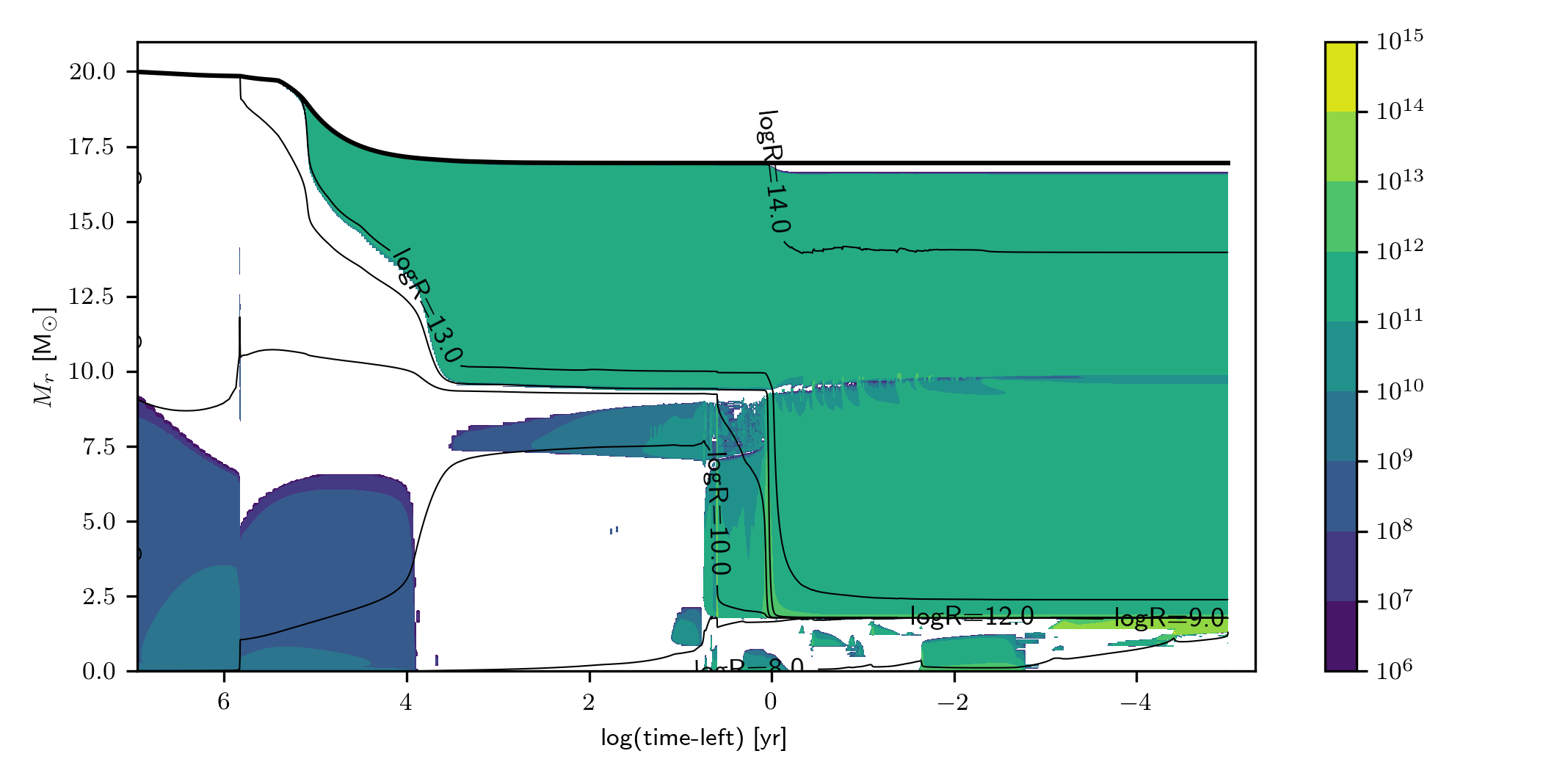}
    \caption{Kippenhahn diagram of the 20SH21 model showing the convective structure using the mass coordinate in solar masses as a function of the Log of time-left before core collapse in years. The colour bar represents the turbulent kinetic energy in convective regions, $\frac{1}{2} v_{\mathrm{conv}}^2$ [cm$^2$] and the isoradius contours show the dramatic expansions caused by the two EBP interactions this model undergoes (the isoradius contours move to lower mass coordinates when expansion occurs). The total mass is shown by the bold black line.}
    \label{fig:2Dmaps20SH21_rad}
\end{figure*}

\begin{figure*}
	\includegraphics[width=\textwidth]{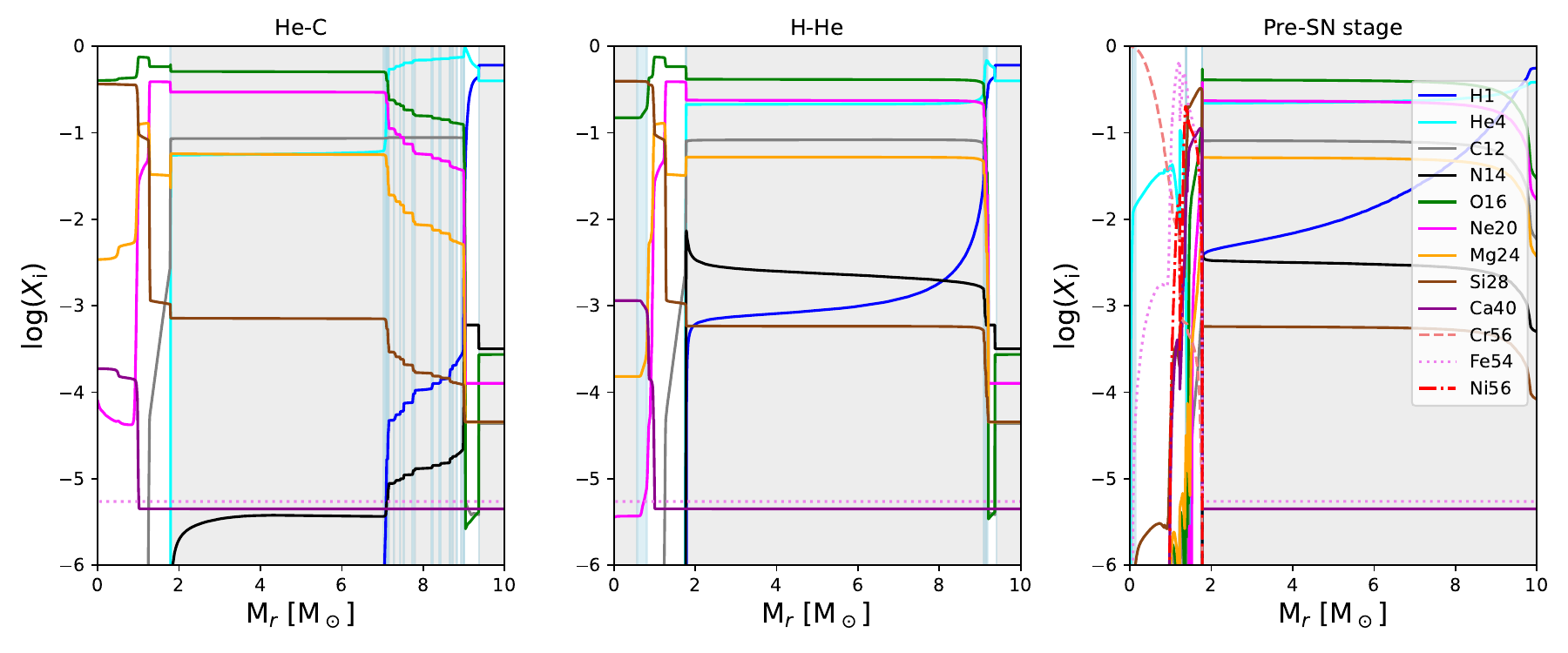}
    \caption{Abundance profiles for the 20SH21 model that undergoes multiple interaction events. The left-hand panel shows the merging of the He and C shells as shown by the alignment of their burning fronts. The entrainment of hydrogen into the He-shell has also started at this point in evolution. This continues to such an extent that the H shell subsequently also merges with the previously merged He-C shell (right-hand panel) to form a merged H-He-C shell at the pre-SN stage, with the mass coordinates $1.8 \lesssim M/\mathrm{M}_{\odot} \lesssim 9$.}
    \label{fig:20SH21abundance}
\end{figure*}

\begin{figure*}
	\includegraphics[width=\textwidth]{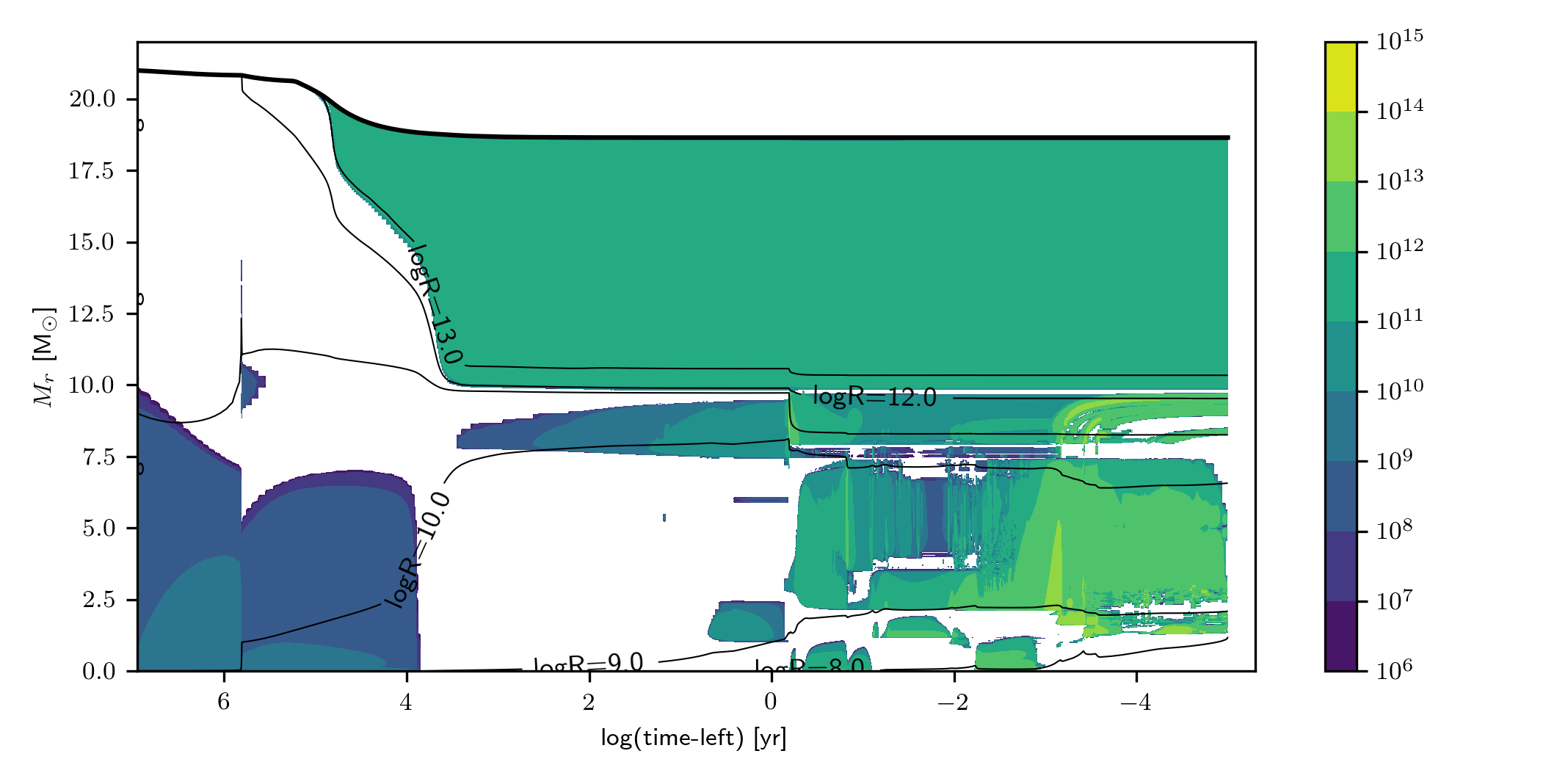}
    \caption{Kippenhahn diagram, as a function of the mass coordinate against the log of the time left before core collapse, for the 21SH21 model that undergoes 4 separate nuclear burning shell interactions. The first is an EBP H-He interaction that causes expansion, shown by the $\mathrm{logR} = 12.0$ isoradius contour, in the mass coordinate range $8 \lesssim M/M_{\odot} \lesssim 9.5$ and a $\mathrm{log(time-left)} \approx 0$. The following three interactions occur after the end of core silicon burning and trigger no significant expansion of the surrounding layers. They are He-C, C-Ne and Si-O interactions respectively. }
    \label{fig:2Dmaps_21SH21}
\end{figure*}

\begin{figure*}
	\includegraphics[width=\textwidth]{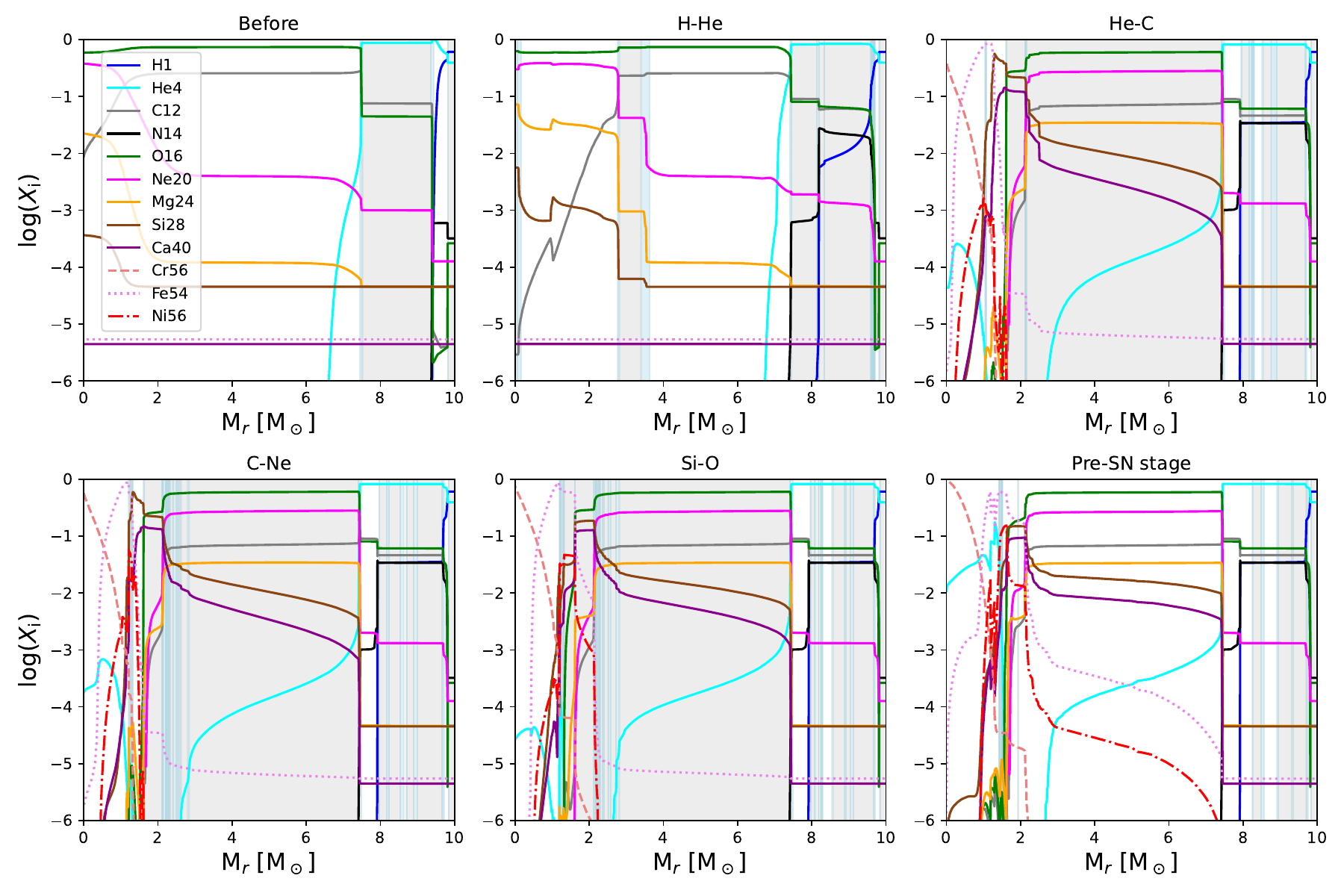}
    \caption{Abundance profiles for the 21SH21 model, showing multiple interaction events. Top row: Before any interactions occur (left), hydrogen being entrained into the He-shell down to a mass coordinate of $\approx 8\mathrm{M}_{\odot} $(middle) and the mixing of helium into the C-shell in the range $2.8 \lesssim M/M{_\odot} \lesssim 7.5$ (right). Bottom row: The merging of the C- and Ne-shells at a mass coordinate of approximately 2M${_\odot}$ (left), and the silicon and oxygen shells also at this approximate location later in evolution (middle). The final abundance profile for this model (right) shows the structure of the 21SH21 at the pre-SN stage. It shows that $^{\mathrm{54}}\mathrm{Fe}$ and $^{\mathrm{56}}\mathrm{Co}$ are mixed to high mass coordinates of around 7.5M${_\odot}$.}
    \label{fig:21SH21abundance}
\end{figure*}

Several models in the two grids experience multiple interaction events throughout their evolution and they generally first undergo an EBP interaction and then a LBP interaction afterwards. This combination of events significantly affects the rest of the star's life. In this Section we discuss the two most extreme examples, 20SH21 and 21SH21, and look at the impact the multiple events have on the further evolution and nucleosynthesis of the star.

\subsubsection{20SH21} \label{20SH21}

The 20SH21 model undergoes two EBP interactions and Fig.\,\ref{fig:2Dmaps20SH21} shows their effects on evolution. At $\mathrm{Log(time-left)} \approx 0.5$, at approximately the same time as the onset of core neon burning, a He-C interaction occurs at a mass coordinate of 2\,M${_\odot}$. The isomass contours show the rapid expansion of the inner 2\,M${_\odot}$ after the interaction,  with the 2\,M${_\odot}$ mass coordinate increasing from a $\mathrm{Log(Radius)} \approx 10$ to $\mathrm{Log(Radius)} \approx 12.5$. The He-C merger is almost immediately followed by an H-He interaction that begins approximately 1 year before the collapse of the core, and triggers a second, more extreme expansion of the stellar interior. The specific kinetic energy of the convective motions in this expanded region increase by an order of magnitude and persists until collapse. Figure \ref{fig:2Dmaps20SH21_rad} shows another version of the Kippenhahn diagram for the 20SH21 model, with the mass coordinate as a function of $\mathrm{Log(time-left)}$ before collapse in years. It shows that the interaction takes place in the mass coordinate range $2 \lesssim M/\mathrm{M}_{\odot} \lesssim 10$. This is also shown in the abundance profiles for the 20SH21 model presented in Figure \ref{fig:20SH21abundance}, where the first panel shows the abundance profile taken just after the helium and carbon shells merge, indicated by the alignment of their respective burning fronts at $\approx 1.8$M$_{\odot}$. The second panel shows that hydrogen is entrained into the newly-formed He-C convective shell to such an extent that it also merges, resulting in an H-He-C convective shell with the mass coordinates $1.8 \lesssim M/M_{\odot} \lesssim 9$. The final panel shows that this new shell remains merged, with hydrogen still being continuously entrained, until the collapse of the core at the pre-SN stage. Examining the effect on the CO-core mass, there is a significant reduction between the end of core He-burning with $M_{\mathrm{CO}} = 7.024\mathrm{M}_\odot$ and the end of evolution where $M_{\mathrm{CO}} = 1.778\mathrm{M}_\odot$. Such a considerable decrease in CO-core mass strongly affects the final fate of the star, as discussed later in Section \ref{FinalFates}.

\subsubsection{21SH21} \label{21SH21}

Another example of a star undergoing multiple interactions is the 21SH21 model. Figure \ref{fig:2Dmaps_21SH21} shows the rapid expansion at $\mathrm{log(time-left)} \approx 0$, approximately the same time as the onset of core neon burning, that is triggered by an EBP interaction where hydrogen is entrained into the convective helium shell. Although there is an increase in kinetic energy immediately after the interaction, it is not sustained before reducing to the pre-interaction level. The abundance profiles in Fig.\,\ref{fig:21SH21abundance} illustrate the series of interactions. The top left panel shows the structure of the star before any interactions occur, the top middle panel shows the H-He interaction discussed above, where hydrogen is entrained down into the helium shell to a mass coordinate $\approx 8$M${_\odot}$. After this, there are three additional nuclear burning shell interactions that occur in this 21SH21 model, all taking place after the end of core silicon burning and within half a day of the collapse of the iron core. The first of these is an He-C interaction that mixes helium into the carbon burning shell down to a mass coordinate $\approx 3$M${_\odot}$ (top right panel). Following this, the carbon and neon burning shells merge, with their burning fronts aligning at $\approx 2$M${_\odot}$ (bottom left panel). Finally, the silicon and oxygen burning shells merge at $\approx 1.75$M${_\odot}$, shown in the bottom middle abundance profile. Interestingly, the final abundance profile (bottom right), taken at the pre-SN stage, shows that the iron group elements are mixed outwards to approximately 7.5M${_\odot}$ due to intermittent interactions between the Si-O and C-Ne merged regions. 


\section{Final Fates} \label{FinalFates}

\begin{table*}
\centering 
\caption{Properties of the SH21 grid at the pre-SN stage (columns 1-14): model codename, CO-core mass ($M_{\mathrm{CO}}$), effective temperature (Log $T_{\mathrm{eff}}$), radius (Log $R$), luminosity (Log $L$), interaction type,  
total final mass ($M_{\mathrm{fin}}$), 
total lifetime ($\tau$), central entropy ($s_c$), mass of the carbon-free core $M_{\mathrm{C-free}}$, compactness parameter measured at a mass coordinate of 2.5\,M${_\odot}$ ($\xi_{\mathrm{2.5}}$) and the mass derivative ($\mu_{\mathrm{4}}$), mass coordinate ($M_{\mathrm{4}}$) and M$_{4}\mu_{4}$, all calculated at the mass coordinate where the entropy per nucleon $s = 4$.}
\label{tab:finalA_table}
    \begin{tabular}{cccccccccccccc}
        \toprule
        Name & $M_\mathrm{CO}$ & Log($T_\mathrm{eff}$) & Log($R$) & Log($L$) & Interactions & $M_{\mathrm{fin}}$ & $\tau$ & $s_c$ & $M_\mathrm{C-free}$  & $\xi_\mathrm{2.5}$ & $\mu_{4}$  & $M_{4}$ & M$_{4}\mu_{4}$ \\
         & [$\mathrm{M}_\odot$] & [$\mathrm{K}$] & [$\mathrm{R}_\odot$] &  [$\mathrm{L}_\odot$] &  & [$\mathrm{M}_\odot$] & [Myr] & $\left[ \frac{k_\mathrm{B}}{\mathrm{baryon}} \right]$ &  [$\mathrm{M}_\odot$] &  & $\left[ \frac{\mathrm{M_\odot}}{1000\,\mathrm{km}} \right]$ & [$\mathrm{M}_\odot$] & $\left[ \frac{\mathrm{M_\odot}^2}{1000\,\mathrm{km}} \right]$  \\
         \midrule
        10SH21 & 2.282 & 3.579 & 2.800 & 4.871 & C-Ne-O & 9.435 & 27.179 & 0.704 & 2.265 & 0.041 & 0.047 & 1.685 & 0.080 \\
        11SH21 & 2.734 & 3.577 & 2.852 & 4.968 & C-Ne-O & 10.196 & 23.526 & 0.757 & 2.615 & 0.084 & 0.050 & 1.552 & 0.078 \\
        12SH21 & 3.222 & 3.577 & 2.894 & 5.050 & C-Ne-O & 11.142 & 20.770 & 0.809 & 3.024 & 0.110 & 0.041 & 1.734 & 0.071 \\
        13SH21 & 3.702 & 3.578 & 2.926 & 5.117 & C-Ne-O & 11.944 & 18.601 & 0.883 & 3.019 & 0.164 & 0.053 & 1.868 & 0.100 \\
        14SH21 & 4.254 & 3.577 & 2.959 & 5.182 & C-Ne-O & 12.174 & 16.870 & 0.971 & 3.157 & 0.247 & 0.069 & 2.032 & 0.140 \\
        15SH21 & 4.787 & 3.581 & 2.978 & 5.236 & C-Ne-O & 13.243 & 15.545 & 1.040 & 3.560 & 0.391 & 0.104 & 2.225 & 0.231 \\
        16SH21 & 5.212 & 3.583 & 2.996 & 5.276 & C-Ne-O & 13.791 & 14.252 & 1.044 & 3.674 & 0.389 & 0.111 & 2.181 & 0.242 \\
        17SH21 & 5.634 & 3.585 & 3.006 & 5.305 & C-Ne-O & 14.862 & 13.180 & 1.032 & 4.032 & 0.376 & 0.103 & 2.189 & 0.226 \\
        18SH21 & 6.026 & 3.587 & 3.018 & 5.338 & C-Ne-O & 15.136 & 12.276 & 0.995 & 4.181 & 0.293 & 0.083 & 2.062 & 0.171 \\
        19SH21 & 6.481 & 3.589 & 3.027 & 5.365 & C-Ne-O & 15.142 & 11.495 & 1.002 & 4.406 & 0.317 & 0.086 & 2.130 & 0.184 \\
        20SH21 & 1.778 & 3.574 & 3.234 & 5.716 & \makecell{He-C\ H-He} & 16.948 & 10.833 & 0.717 & 1.778 & $\sim 0$\footnote{exact value is 2.3$\times 10^{-5}$} & 0.040 & 1.413 & 0.056 \\
        21SH21 & 7.417 & 3.589 & 3.082 & 5.475 & Multiple & 18.644 & 10.246 & 0.763 & 2.129 & 0.135 & 0.069 & 1.620 & 0.111 \\
        22SH21 & 8.005 & 3.589 & 3.169 & 5.649 & \makecell{Multiple} & 16.883 & 9.731 & 0.860 & 2.252 & 0.177 & 0.082 & 1.690 & 0.139 \\
        23SH21 & 8.373 & 3.597 & 3.171 & 5.683 & \makecell{Multiple} & 19.725 & 9.286 & 0.991 & 2.659 & 0.301 & 0.096 & 1.943 & 0.186 \\
        24SH21 & 9.068 & 3.597 & 3.102 & 5.543 & \makecell{Multiple} & 18.190 & 8.875 & 0.813 & 4.066 & 0.208 & 0.073 & 1.686 & 0.122 \\
        25SH21 & 9.383 & 3.598 & 3.121 & 5.586 & \makecell{H-He\ He-C\ C-Ne} & 18.256 & 8.521 & 0.961 & 2.313 & 0.279 & 0.101 & 1.915 & 0.194 \\
        30SH21 & 3.162 & 3.601 & 3.158 & 5.673 & \makecell{He-C\ C-Ne-O} & 22.977 & 7.182 & 0.803 & 2.513 & 0.130 & 0.066 & 1.582 & 0.104 \\
        35SH21 & 2.714 & 3.621 & 3.175 & 5.786 & \makecell{He-C\ C-Ne-O-Si} & 19.570 & 6.346 & 0.701 & 2.635 & 0.083 & 0.040 & 1.473 & 0.058 \\
        40SH21 & 4.589 & 3.623 & 3.220 & 5.885 & He-C & 23.085 & 5.715 & 0.989 & 2.912 & 0.338 & 0.126 & 2.000 & 0.251 \\
        45SH21 & 5.351 & 3.626 & 3.264 & 5.984 & He-C & 26.039 & 5.248 & 1.078 & 3.737 & 0.464 & 0.148 & 2.190 & 0.323 \\
        \bottomrule
    \end{tabular}
\end{table*}

\begin{table*}
\centering
\caption{Same as Table\,\ref{tab:finalA2_table} for the f0p02 grid at the pre-SN stage. }
\label{tab:finalA2_table}
    \begin{tabular}{cccccccccccccc}
        \toprule
        Name & $M_\mathrm{CO}$ & Log($T_\mathrm{eff}$) & Log($R$) & Log($L$) & Interactions & $M_{\mathrm{fin}}$ & $\tau$ & $s_c$ & $M_\mathrm{C-free}$  & $\xi_\mathrm{2.5}$ & $\mu_{4}$  & $M_{4}$ & M$_{4}\mu_{4}$ \\
         & [$\mathrm{M}_\odot$] & [$\mathrm{K}$] & [$\mathrm{R}_\odot$] &  [$\mathrm{L}_\odot$] &  & [$\mathrm{M}_\odot$] & [Myr] & $\left[ \frac{k_\mathrm{B}}{\mathrm{baryon}} \right]$ &  [$\mathrm{M}_\odot$] &  & $\left[ \frac{\mathrm{M_\odot}}{1000\,\mathrm{km}} \right]$ & [$\mathrm{M}_\odot$] & $\left[ \frac{\mathrm{M_\odot}^2}{1000\,\mathrm{km}} \right]$ \\
        \midrule
        10f0p02 & 1.942 & 3.587 & 2.735 & 4.772 & C-Ne-O & 9.752 & 25.615 & 0.769 & 1.904 & 0.010 & 0.018 & 1.632 & 0.030 \\
        11f0p02 & 2.198 & 3.587 & 2.775 & 4.852 & C-Ne-O & 10.780 & 21.963 & 0.768 & 2.165 & 0.032 & 0.047 & 1.572 & 0.073 \\
        12f0p02 & 2.482 & 3.588 & 2.809 & 4.922 & C-Ne-O & 11.790 & 19.229 & 0.712 & 2.442 & 0.082 & 0.069 & 1.467 & 0.101 \\
        13f0p02 & 2.808 & 3.589 & 2.837 & 4.983 & C-Ne-O & 12.754 & 17.183 & 0.727 & 2.756 & 0.131 & 0.062 & 1.695 & 0.106 \\
        14f0p02 & 3.082 & 3.591 & 2.856 & 5.030 & C-Ne-O & 13.674 & 15.524 & 0.762 & 2.986 & 0.171 & 0.075 & 1.743 & 0.131 \\
        15f0p02 & 3.321 & 3.593 & 2.872 & 5.070 & C-Ne-O & 14.769 & 14.157 & 0.790 & 3.219 & 0.156 & 0.063 & 1.677 & 0.105 \\
        16f0p02 & 3.657 & 3.594 & 2.894 & 5.118 & C-Ne-O & 15.742 & 13.017 & 0.881 & 3.408 & 0.204 & 0.080 & 1.821 & 0.145 \\
        17f0p02 & 3.969 & 3.595 & 2.912 & 5.160 & C-Ne-O & 16.712 & 12.062 & 0.923 & 3.707 & 0.229 & 0.079 & 1.901 & 0.150 \\
        18f0p02 & 4.315 & 3.596 & 2.934 & 5.205 & C-Ne-O & 17.672 & 11.248 & 0.966 & 4.058 & 0.285 & 0.086 & 2.025 & 0.174 \\
        19f0p02 & 4.778 & 3.596 & 2.953 & 5.244 & C-Ne-O & 18.591 & 10.552 & 1.029 & 3.835 & 0.383 & 0.103 & 2.207 & 0.227 \\
        20f0p02 & 5.097 & 3.597 & 2.971 & 5.283 & C-Ne-O & 19.568 & 9.962 & 1.053 & 4.413 & 0.422 & 0.108 & 2.273 & 0.246 \\
        21f0p02 & 5.395 & 3.600 & 2.963 & 5.281 & C-Ne-O & 20.531 & 9.443 & 1.075 & 4.584 & 0.481 & 0.126 & 2.302 & 0.290 \\
        22f0p02 & 5.808 & 3.610 & 2.898 & 5.192 & C-Ne-O & 21.459 & 8.982 & 0.869 & 2.323 & 0.214 & 0.073 & 1.771 & 0.129 \\
        23f0p02 & 6.182 & 3.612 & 2.914 & 5.230 & C-Ne-O & 22.408 & 8.580 & 0.815 & 2.223 & 0.224 & 0.080 & 1.854 & 0.148 \\
        24f0p02 & 6.567 & 3.610 & 2.938 & 5.271 & C-Ne-O & 23.322 & 8.220 & 0.852 & 3.784 & 0.206 & 0.094 & 1.690 & 0.158 \\
        25f0p02 & 7.020 & 3.608 & 2.977 & 5.340 & C-Ne-O & 24.209 & 7.895 & 1.074 & 6.730 & 0.464 & 0.113 & 2.316 & 0.262 \\
        30f0p02 & 9.068 & 3.608 & 3.068 & 5.521 & C-Ne-O & 28.459 & 6.690 & 1.085 & 6.788 & 0.518 & 0.225 & 1.881 & 0.424 \\
        35f0p02 & 10.893 & 3.607 & 3.243 & 5.867 & \makecell{H-He\ C-Ne-O} & 33.154 & 5.904 & 0.929 & 2.238 & 0.215 & 0.109 & 1.769 & 0.192 \\
        40f0p02 & 13.276 & 3.618 & 3.156 & 5.735 & \makecell{He-C\ C-Ne-O} & 32.916 & 5.356 & 1.101 & 2.950 & 0.543 & 0.241 & 2.033 & 0.491 \\
        45f0p02 & 7.364 & 3.515 & 3.359 & 5.733 & \makecell{H-He\ He-C} & 35.255 & 4.947 & 1.065 & 2.929 & 0.445 & 0.138 & 2.160 & 0.299 \\
        \bottomrule
    \end{tabular}
\end{table*}

One of the key goals of stellar evolution modelling is to predict the final fate of stars as it is rare to be able to observe SN progenitors and thus hard to connect a fate with the initial properties of the star purely from observations. Ideally, one would simulate the entire evolution and the core collapse (and the possible explosion) using three-dimensional (3D) (magneto-)hydrodynamical simulations. While it is finally possible to simulate the core collapse in 3D \citep{Burrows2019, Muller2019, Muller2020,Sandoval2021,Burrows2024,Janka2025}
as well as study the effects of magnetic fields during the collapse \citep{Mosta2015, Obergaulinger2021, MullerVarma2020, Muller2024}, it will never be possible to simulate the entire evolution of stars in 3D due to the very disparate time scales between convective turnover time scales and the lifetime of stars. This motivates the 321D approach \citep{Arnett2015} developing synergies between expensive 3D simulations testing and guiding 1D models and these 1D models that can follow the entire structure and evolution of stars. Using progenitor structures like the one computed in this study, there are various options to estimate their fate. Future studies will explode our progenitors in multi-D and we have developed a method to initialise these multi-D simulations with 2D \citep{Muller2015} and now 3D synthetic velocity fields for convective regions (Varma et al, in prep.), which will help determine the impact of progenitor asymmetries on the fate of massive stars. In this study, however, we want to analyse the pre-SN structure to assess the impact of the initial-mass dependent SH21 CBM on the fate as well as compare our models to the literature. For this, several methods have been developed and used in the literature including compactness \citep{Ott2011}, the Ertl 2-parameter criterion \citep{Ertl2016}, central entropy \citep{Schneider2024} or analytical methods to simplify the core-collapse simulation \citep{BM2016}.

\subsection{The Evolution of the CO-core Mass} \label{CO-core-evol}

\begin{figure*}
	\includegraphics[width=\textwidth]{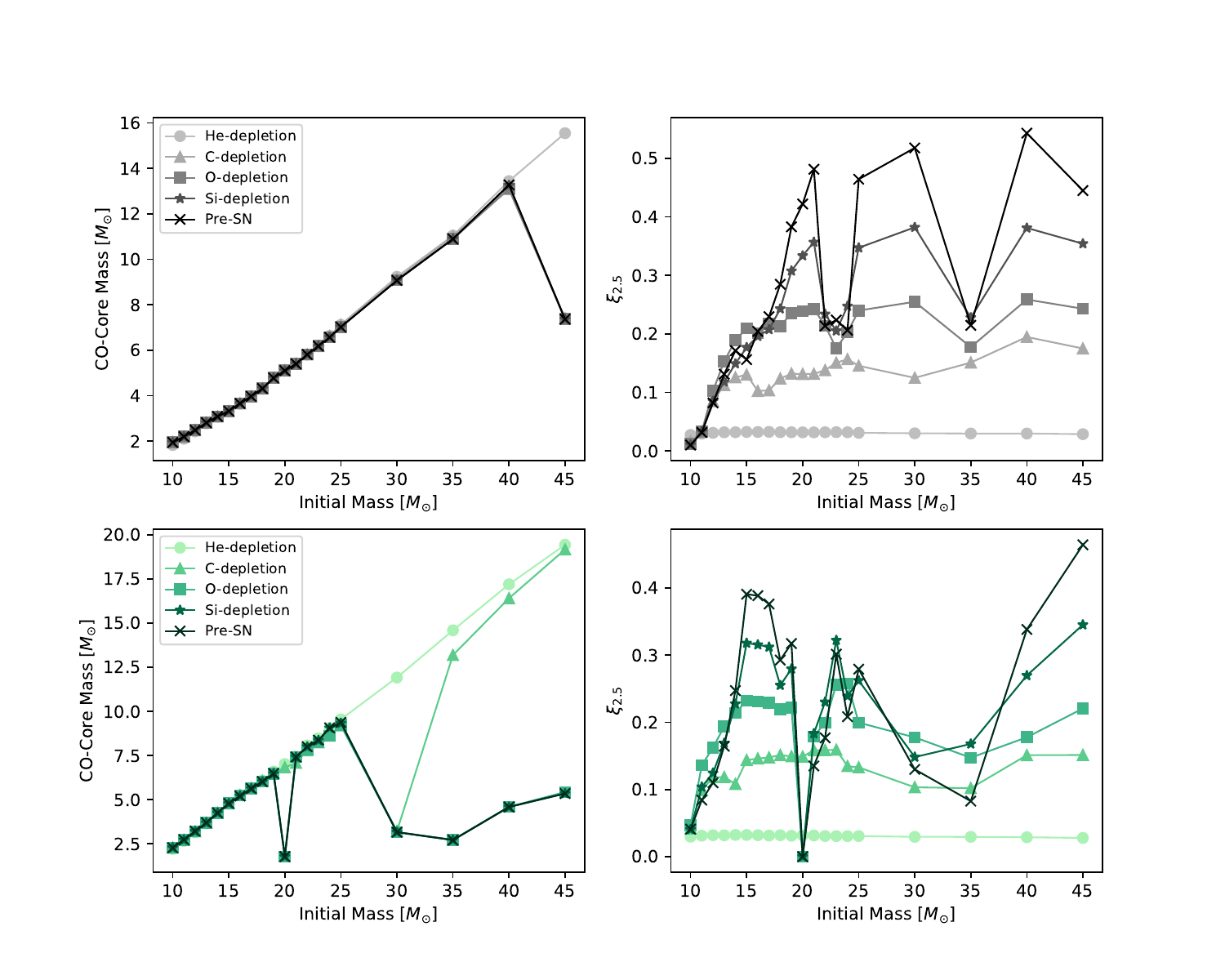}
    \caption{Comparison of the CO-core mass (left) and the compactness parameter $\xi_{\mathrm{2.5}}$ (right) at the end of each stage of core nuclear burning, in addition to at the end of evolution (pre-SN), as a function of initial mass. The top row refers to the f0p02 grid and the bottom row to the SH21 grid. In both cases, darker colours represent later evolutionary stages.}  
    \label{fig:Compactness_stage_comparison}
\end{figure*}

\begin{figure*}
	\includegraphics[width=\textwidth]{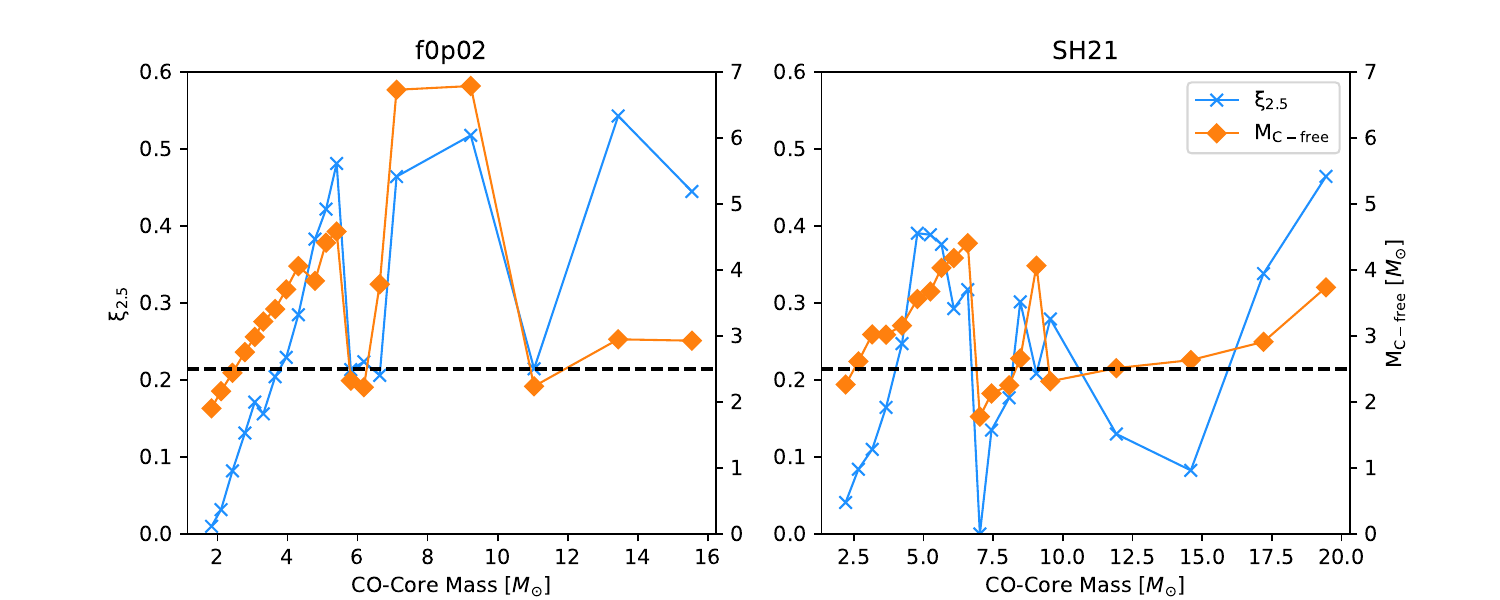}
    \caption{Comparison of the compactness parameter ($\mathrm{\xi_{2.5}}$) and the mass of the carbon-free core ($\mathrm{M_{C-free}}$), both taken at the pre-SN stage, against the CO-core mass for both grids of models. The black dashed line represents the mass coordinate at which the compactness is evaluated (2.5M${_\odot}$).}
    \label{fig:Compactness_c_free}
\end{figure*}

\begin{figure*}
	\includegraphics[width=\textwidth]{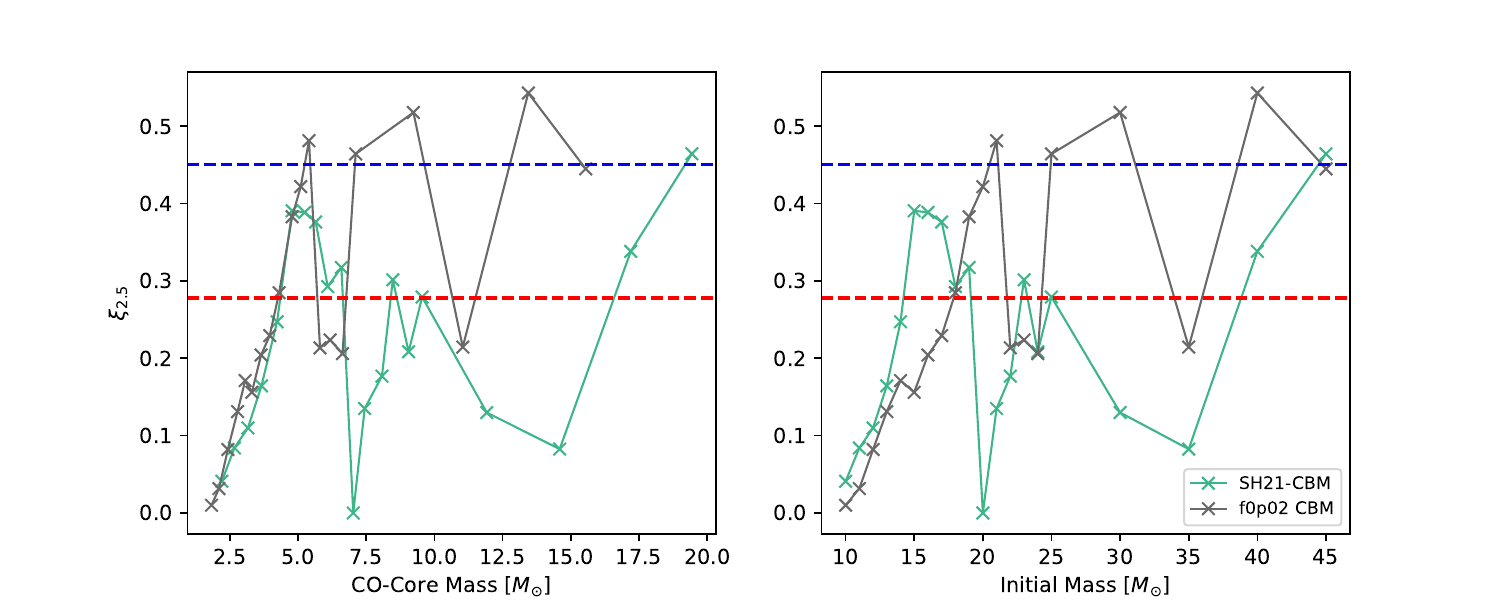}
    \caption{Compactness parameter measured at 2.5M${_\odot}$ as a function of the mass of the CO-core at the end of core He-burning (left) and of initial mass for reference (right). The dashed red line corresponds to the compactness value at which \citet{BM2016} suggests the predicted fates switch from neutron star to black hole formation ($\xi_{\mathrm{2.5}} = 0.278$), whereas the blue dashed line refers to the value predicted by \citet{Ott2011} ($\xi_{\mathrm{2.5}} = 0.45$).}
    \label{fig:Compactness_Subplot}
\end{figure*}

While it would make sense to analyse the pre-SN structure directly as a function of the initial mass, several studies have shown that a more practical approach is the use the CO-core mass ($M{_\mathrm{CO}}$, defined in this work as the mass coordinate where the helium abundance first falls below 1\% starting from the surface) as an intermediary. We can then study the relation between the initial mass and $M{_\mathrm{CO}}$ first to understand the role that processes like mass loss, binary interactions and mixing during the early phases have on the evolution of massive stars. The relation between $M{_\mathrm{CO}}$ at the end of core He-burning and the initial mass is presented in Fig.\,\ref{fig:MCOvMini} for both grids and was already discussed in Section\,\ref{Surface&Central}. However, throughout a star's evolution, its CO-core mass may change as a result of events that affect its internal structure so it is useful to review how it evolves with time. 
Generally, the CO-core mass for each model is established at the end of core He-burning and increases monotonically with initial mass. This can be seen in the top left panel of Fig.\,\ref{fig:Compactness_stage_comparison}, which shows the mass of the CO-core at the end of each core burning stage against initial mass for the f0p02 grid. The end of each burning phase is defined as the point in evolution at which the central abundance of that isotope first drops below 1\%. However, the bottom left panel shows how interactions occurring in the SH21 grid can affect CO-core mass throughout the evolution. For example, the diving He-shells and He-C interactions experienced by SH21 models with M$_{\mathrm{initial}} \geq 35$M${_\odot}$ cause a significant reduction in the CO-core mass after the end of core carbon burning. This makes sense as interactions involving the He-burning shell will modify the abundance profile of helium, which is used to determine the CO-core mass.  
The reduction in $M{_\mathrm{CO}}$ seen in the 30SH21 and 45f0p02 models are also caused by diving He-shells and He-C interactions, although these occur slightly earlier in these models, before the end of core carbon burning. These slightly earlier interactions cause a less severe decrease in CO-core mass in the 45f0p02 model ($M{_\mathrm{CO}}$ at the end of evolution, column 2 in Tables \ref{tab:finalA_table} and \ref{tab:finalA2_table}, is $\approx 47\%$ that at the end of core He-burning, last column in Tables \ref{tab:earlyA_table} and \ref{tab:earlyA2_table}) compared to the later interactions experienced by the 45SH21 model (final $M{_\mathrm{CO}} \approx 27\%$ that at core He-depletion). The two EBP interactions that occur in the 20SH21 model can also been seen to cause a significant drop in the $M{_\mathrm{CO}}$ after the end of core carbon burning. As a result of the effects of interactions on M${_\mathrm{CO}}$ later in evolution, we use the CO-core mass calculated at the end of core He-burning to compare against for the predictions of the final fates of the models presented in this work. This is also useful when comparing to other studies which rarely have shell interactions modifying the CO-core mass late in the evolution or in which models may not have reached core collapse. It also maintain the main goal of the use of the CO-core mass, which is to take care of the dependence on the early evolution and in particular the effect of mass loss and binary interactions.

\subsection{Explodability Criteria} \label{Explodability} 

The final fate of stars are uncertain and very difficult to predict. However, as discussed above, methods of making these predictions using stellar and supernova theory and observations have been developed to try and predict the likelihood of a progenitor successfully exploding as a supernova, i.e. a progenitor's `explodability'. The fate predictions are important as understanding the likely fates of different types of stars will, in turn, aid understanding of the nucleosynthetic processes occurring during the very end of a star's life and its impact on GCE.

\subsubsection{Compactness Parameter and the Explodability of Models} \label{Compactness}

One of the commonly used measure of the explodability of stars is the compactness parameter that was first introduced by \citet{Ott2011}, where the outcomes of unsuccessful core collapse supernovae (CCSNe) and black hole formation were studied for single massive stars. They found that, for a certain equation of state (EOS), they can predict the formation of black holes based only on a single parameter; the compactness of the stellar model at the pre-SN stage. Defined by the equation,

\begin{equation} \label{compactness_eqn}
\xi_{\mathrm{2.5}} = \frac{M_{\mathrm{2.5}\,M_{\odot}}/\mathrm{M}_{\odot}}{R_{\mathrm{2.5\,M_{\odot}}}/\mathrm{1000\,km}} 
\end{equation}

the compactness parameter $\xi_{2.5}$ is a dimensionless parameter used as a measure of a star's explodability. It is the ratio of the innermost 2.5M${_\odot}$ and the radius of this mass coordinate ($R_{\mathrm{2.5\,M_{\odot}}}$). Note that we calculated the compactness parameter for different mass coordinates and assess its sensitivity to that choice in Appendix \ref{Appendix_compactness}. Although it is usually measured at the very end of evolution, at the pre-SN stage, highlighting how $\xi_{2.5}$ changes with progressive core nuclear burning stages demonstrates the effect of nuclear burning interactions on the internal structure. This is because some models that undergo interactions do not show the expected continual increase in the compactness parameter throughout their evolution, depending on the mass-coordinate that the interaction(s) occur at. This is shown in the right-hand panels of Figure \ref{fig:Compactness_stage_comparison}, where the compactness parameter is taken at the end of each core burning stage as a function of initial mass for both grids of models: f0p02 grid in the top panel (grey) and SH21 in the bottom panel (green).
As for the evolution of the CO-core mass, the f0p02 models generally show the common trend seen published models with  $\xi_{2.5}$ gradually increasing during the evolution and peaks getting more and more pronounced up to the pre-SN stage.
The situation is quite different for the SH21 models. For example, in the 12SH21 model, $\xi_{2.5}$ increases until the end of core oxygen burning, after which point the LBP C-Ne-O interaction occurs at a mass coordinate of $\approx 1.75$M${_\odot}$ (for the bottom of the original oxygen shell) and the compactness drops. This drop is due to the expansion of the merged shells caused by the increased energy generated by carbon (and neon) reaching the bottom of the oxygen burning front.
The 20SH21 model also shows a very sharp drop in compactness after the end of core carbon burning as a result of the expansion caused by the multiple EBP interactions that it undergoes. The 21SH21 model, which contains 4 interactions, also shows a drop in the compactness after the end of core silicon burning. Overall, if an interaction occurs and causes an expansion of the layers at or within the 2.5M${_\odot}$ mass-coordinate, then the compactness parameter will be affected, with stronger interactions causing a larger drop.

After reviewing the evolution of the CO-core mass and $\xi_{2.5}$, we can now discuss how they relate to each other in Fig. \ref{fig:Compactness_c_free} and compare our models to other studies.
Up to a CO-core mass of about 5.5\,M${_\odot}$ ($M_\mathrm{ini}=21$\,M${_\odot}$) for the f0p02 grid, compactness increases gradually up to a peak around 0.5. This is then followed by a narrow sharp drop to $\xi_{2.5}=0.2$ linked to the change from convective to radiative core carbon burning (see Kippenhahn diagrams in the Supplementary material). The drop continues only up to a CO-core mass of 6.6\,M${_\odot}$, $M_\mathrm{ini}=24$\,M${_\odot}$ where the next broad peak around 0.5 begins (with the exception of the $M_\mathrm{ini}=35$\,M${_\odot}$ model).
While the sudden drop caused by the transition from convective to radiative core carbon burning has already been observed and discussed in many studies \citep{ChieffiLimongi2020, Sukhbold2014, Sukhbold2016, Sukhbold2018}, a notable difference in the f0p02 grid compared to the literature is the gradual increase of $\xi_{2.5}$ with initial/CO-core mass in the lower mass range. Indeed, other studies generally finds an initial plateau at low $\xi_{2.5}$ values \citep[see e.g. Figures 4 and 5 in][]{Temaj2024}, followed by a rise to the peak \citep[see also Figure 1 in][]{Laplace2025}. This difference is likely due to the fact we include CBM at all convective boundaries (as 3D simulations suggest is the case) rather just for the top of the H- and He-burning convective cores.

The pre-SN compactness parameter generally behaves similarly in the SH21 grid for the lower mass range. $\xi_{2.5}$ increases gradually up until the first compactness peak, although the peak shifts to a lower initial mass than in the f0p02 grid; at 16M${_\odot}$ in the SH21 grid. This is a result of the more efficient CBM mixing more fuel into the burning regions and leading to longer nuclear burning phases, larger core masses and ultimately more compact cores at lower initial masses. The first peak is reached at a similar CO-core mass in both grids as shown in Fig.\,\ref{fig:Compactness_Subplot}, showing again that pre-SN properties are much similar when we plot them as a function of the CO-core mass rather than the initial mass. From the first peak onward, the many shell interactions lead to a less clear picture and generally lower compactness than for the corresponding f0p02 models (left panel in Fig.\,\ref{fig:Compactness_Subplot}). 

In Fig.\,\ref{fig:Compactness_c_free}, we plot another quantity that may relate or explain the compactness of the pre-SN models: the mass of the carbon-free core (M${_\mathrm{C-free}}$, right-hand-side axis), which corresponds to the bottom of the convective region where carbon gets completely depleted. This is located at the bottom of the carbon burning shell unless a shell interaction brings carbon down to the Ne- or O-burning shells. Such interactions lead to enhanced energy generation and expansion of the layers involved, reducing their compactness. It is thus not a surprise that the patterns for M${_\mathrm{C-free}}$  closely correlate with the compactness parameter when both are measured at the pre-SN stage. In Fig.\,\ref{fig:Compactness_c_free}, the dashed black horizontal line depicts the mass-coordinate where the compactness is evaluated (2.5M${_\odot}$. If the M${_\mathrm{C-free}} < 2.5\mathrm{M}_{_\odot}$, the compactness will be very low. This is because M${_\mathrm{C-free}}$ shows the location of the carbon-burning front at the end of evolution and the lower the mass coordinate of this burning front, the more extended the carbon-burning shell and thus the less compact the layers below. Therefore, if the interactions lower the M${_\mathrm{C-free}}$ to below 2.5M${_\odot}$, then that model will have a low compactness. Note that for some models, 45f0p02 for example, the compactness strongly correlates to the mass coordinate it is evaluated at (see Appendix \ref{Appendix_compactness}). The right-hand panel of Figure \ref{fig:Compactness_c_free} shows that, whilst there is still a correlation between the M${_\mathrm{C-free}}$ and the compactness parameter for the SH21 grid, their values are generally lower as a result of the interactions they undergo. This is especially apparent in the models with $ 7 \lesssim \mathrm{M_{CO}}/\mathrm{M_{\odot}} \lesssim 9.5 $ (corresponds to models 20SH21 to 25SH21) where multiple interactions lower the M${_\mathrm{C-free}}$ and therefore also the compactness parameter.

Figure \ref{fig:Compactness_Subplot} shows the compactness parameter, taken at the pre-SN stage, for each model in the two grids as a function of CO-core mass (left) and initial mass (right). According to \citet{Ott2011}, any star with a pre-SN compactness parameter below 0.45 ($\xi_{\mathrm{O11}}$ hereinafter) is predicted to explode as a successful supernova. Whilst if the compactness is above 0.45, a more probable outcome is that the supernova will fail and will collapse into a black hole. The blue dashed line in Figure \ref{fig:Compactness_Subplot} represents this bifurcation value and it shows that only the 45M${_\odot}$ model in the SH21 grid is predicted to fail to explode compared to the 21, 25, 30 and 45M${_\odot}$ models in the f0p02 grid. However, this bifurcation value is still debated. For example, \citet{Ugliano2012} argues against the idea of a single value and instead introduces the idea of a range of compactness values at which, based on their study of 1D solar metallicity stellar models, either a successful or unsuccessful supernova will occur ($0.15 \lesssim \xi_{2.5} \lesssim 0.35$). There are also initial mass ranges in which \citet{Ugliano2012} predict unsuccessful supernovae. \citet{BM2016} also find a `transition region' between predicted successful and unsuccessful supernova explosions, although their proposed values are higher than those suggested by \citet{Ugliano2012} at $0.2 \lesssim \xi_{2.5} \lesssim 0.4$. Furthermore, \citet{BM2016} also identify a critical value for the compactness parameter of 0.278 ($\xi_{\mathrm{M16}}$ hereinafter), at which they predict the fate will switch from successful to unsuccessful supernovae and eventual black hole formation. This bifurcation value is shown in Figure \ref{fig:Compactness_Subplot} as the red dashed line and it can be seen that the values suggested by \citet{Ott2011} and \citet{BM2016} ($\xi_{\mathrm{O11}}$ and $\xi_{\mathrm{M16}}$ respectively) lead to substantial differences in their predictions of stellar fates in both grids of models. Differences in bifurcation values between studies arise due to the different sets of progenitors and methods used to calculate them. $\xi_{\mathrm{M16}}$ predicts that many more models throughout the initial mass range presented in this work will fail to explode as successful supernovae, with the predicted fates of the SH21 grid being the most affected by the difference in bifurcation values.

\subsubsection{Other Explodability Assessment Methods}
There are other methods of predicting explodability, one being the 2-parameter criterion described in \citet{Ertl2016}. Measured at the pre-SN stage it uses the mass coordinate and mass derivative ($M_{\mathrm{4}}$ and $\mu_{\mathrm{4}}$ respectively), both taken at an entropy per nucleon of 4, to predict black hole or neutron star formation. The mass derivative is given by

\begin{equation} \label{mu4_eqn}
\mu_{\mathrm{4}} = \left. \frac{dm/M_{\odot}}{dr/1000\mathrm{km}} \right\vert\ _{s=4} 
\end{equation}

and using $dm = 0.3\,\mathrm{M}_{\odot}$. The parameters $\mu_{\mathrm{4}}$ and $M_{\mathrm{4}}\mu_{\mathrm{4}}$ are used as proxies for the mass accretion rate $\dot{M}$ and the neutrino luminosity $L_{\mathrm{\nu}}$ respectively and their values are shown in Tables \ref{tab:finalA_table} and \ref{tab:finalA2_table}. In the neutrino-driven mechanism for supernova, a higher mass accretion rate will stall the shock whereas a higher neutrino luminosity will overcome gravity and lead to a supernova explosion. 
Both quantities are taken at an entropy per nucleon of 4 to align with the silicon-oxygen interface and the corresponding density discontinuity. Models that lie below a separation line are predicted to explode as supernovae whereas those that lie above are expected to collapse into black holes. However, the Ertl criterion predictive power is limited by 
the fact that these separation curves are calibrated against specific sets of progenitors. This means that it is unknown if these separation lines can be used when making explodability predictions for models that are not similar to those used to calibrate those separation lines. Furthermore, the Ertl criterion is also limited by the input physics of the models used in the calibration of the separation lines. For example, differences in the nuclear network or the treatment of mass loss or convective boundary mixing between the calibration models and those whose fate is being predicted. Using this method was inconclusive for our set of models as the models predicted to explode from their compactness had no correlation with how likely they were to explode from the Ertl criterion point of view. Furthermore, models did not cluster in the Ertl 2-parameter space as expected, and instead populated a narrow band parallel to the separation line calibrated in past studies. We also provide the central entropy in Tables\,\ref{tab:finalA_table} and \ref{tab:finalA2_table} so that our models can be compared to \cite{Schneider2025} and future studies using this fate predictor.

Another method of predicting the remnant type of a progenitor is by using the SN-analytical method introduced by \citet{BM2016}. This model first decides whether and when the shock is revived using semi-empirical relations for the neutrino heating conditions, which are motivated by multi-dimensional supernova simulations. If shock revival occurs, it uses a set of simple ODEs to estimate the explosion energy and the mass of the neutron star (or a black hole formed by fallback), accounting for energy input by neutrino heating and nuclear burning. One limitation of this SN-analytical method is the number of explicit free parameters (six) that need to be calibrated against multi-dimensional simulations, and whose uncertainties would need to be propagated into the predicted outcomes. This model yields similar results to not only the Ertl criterion, but also the single parameter compactness criterion, which is discussed in Section \ref{Compactness}.

A further method of predicting progenitor explodability was proposed by \citet{Wang2022}, in which they derived a condition for explodability based on the final density profiles of progenitor models, using the steepness and time of accretion of the density discontinuity at the Si/O interface. As the location of this density discontinuity is accreted, the ram pressure ($P_{\mathrm{ram}}$) decreases suddenly, causing a `surge'/expansion \citep{Boccioli2023} that can reinvigorate the shock and lead to a successful CCSNe. By comparing to 2D simulations of the collapse of the progenitor models, \citet{Wang2022} finds that a successful explosion will occur if $\mathrm{max}\left(\frac{\Delta P_{\mathrm{ram}}}{P_{\mathrm{ram}}} \right) > 0.28$, where $\Delta P_{\mathrm{ram}}$ is the difference in ram pressure across the density discontinuity. \citet{Boccioli2023} finds a similar condition, although they derive it using the size of this density jump, $\delta \rho$. They use 1D collapse models to conclude that a progenitor will explode if $\delta\rho_{*}^2/\rho_{*}^{2} > 0.08$, where $\rho_{*}$ is the density as measured at the side of the discontinuity closest to the stellar centre. Making use of the fact that $\delta P_{\mathrm{ram}}/P_{\mathrm{ram}} \approx \delta \rho / \rho$, \citet{Boccioli2023} translates the criterion proposed by \citet{Wang2022} to $\mathrm{max}(\delta \rho^{2}/\rho^{2}) > 0.078$. This shows very good agreement between the two studies, suggesting that density profiles taken a the pre-SN stage of massive stars can be used to predict the explodability of massive stars.

\section{Conclusions} \label{Conclusions}

In this paper we have studied how using an initial mass dependent CBM prescription guided by the results of \citet{Scott2021} and 3D simulations influences (1) the occurrence and type of nuclear burning shell interactions and (2) the impact of the CBM and shell interactions on the fates of  stars at a low initial metallicity, $Z=0.001$. 

One of the main consequences of applying the higher strength SH21 CBM to stellar models is that they behave like stars of higher initial masses but with weaker CBM (f0p02 comparison grid). Furthermore, SH21 CBM also causes longer evolutionary phases, which therefore leads to more massive CO-cores as well as increased luminosities and  MS width. 

The SH21 CBM models (and to a lesser extent the f0p02) experience two main types of nuclear burning shell interactions; early burning phase (EBP) and late burning phase (LBP) interactions. H-He interactions are one type of EBP interaction and occur in the SH21 models with initial masses $20 \leq M/\mathrm{M_{\odot}} \leq 25$ and the 35M${_\odot}$ and 45M${_\odot}$ f0p02 models. They involve hydrogen being entrained into the helium shell, which provides conditions for $i$-process nucleosynthesis. While H-He interactions have been commonly present in Pop III and very low Z massive stellar models \citep{Limongi2012,Clarkson2018, Clarkson2018erratum, Clarkson2021}, the fact that they occur in our models at  $Z=0.001$ using higher CBM likely implies H-He interactions to be more frequent (possibly very frequent) in the early Universe than previously thought. Furthermore, these interactions can be so energetic that they cause dramatic expansion, which in some cases propagates to the surface and increases the radius and luminosity of the star and thus explain some SN precursors. 
Another type of EBP interaction is that between the helium and carbon shells (He-C). They are common in SH21 models with initial masses $\geq 20\mathrm{M}_\odot$ and are also present in the 40f0p02 and 45f0p02 models. They usually accompany other types of interaction and, in models with initial masses $\geq 30\mathrm{M}_\odot$, they follow a diving He-shell. Such diving He-shells and He-C interactions often cause a significant reduction in the CO-core mass, which in turn likely affect their fate. 

LBP interactions are present in the vast majority of models in the two grids and usually consist of a merger between the C-, Ne- and O-shells late in evolution. {These likely occur more often in our models because, unlike many published studies, CBM is applied to every convective boundary throughout evolution in this work.} As described in Section \ref{LPI}, C-Ne-O interactions can have interesting effects on nucleosynthesis as \citet{Roberti2025} show that such mergers could explain abundances for elements like potassium and scandium as well as initiate the $\gamma$-process before the star explodes \citep{Roberti2025l}. 

In these grids of models we also find that several models undergo multiple interactions throughout their evolution (see Tables \ref{tab:finalA_table} and \ref{tab:finalA2_table}). The 20SH21 and 21SH21 models are two examples with the more dramatic consequences. Section \ref{20SH21} describes how the two EBP interactions cause significant expansion of the layers at and above the interaction site, causing the largest increase in luminosity out of the models containing EBP interactions. Meanwhile the 21SH21 model (Section \ref{21SH21}) contains four separate instances of interactions between nuclear burning shells; 1 EBP and 3 LBP interactions. Iron-peak elements are also mixed upward at the end of evolution to a mass coordinate of 7.5M${_\odot}$. 

Finally, in order to investigate the impact of these nuclear burning shell interactions on pre-SN structure and the predicted fates, we calculate the compactness parameter for the models in each grid. We first examine the compactness of each model at the end of each core burning phase, noting that changes to internal structure caused by interactions affect the compactness value, potentially having an effect on the predicted fate of the model. Furthermore, we note that the mass of the carbon-free core (M$_\mathrm{C-free}$) is a good indicator of the compactness parameter when both are measured at the pre-SN stage. In regards to comparing the pre-SN compactness values between the two grids, we find that the compactness peak moves to lower initial masses in the SH21 grid due to their larger cores as a result of the higher CBM used in these models. As a function of the CO-core mass, on the other hands, the gradual rise to the first compactness peak is similar between our two grids of models (while different from past studies most likely due to our models including CBM at all convective boundaries as suggested by 3D simulations). Above the first peak, the f0p02 grid models show a narrow drop followed by a second broad compactness peak starting around 25\,M$_\odot$ as found in previous studies, while the SH21 models remain less compact due to multiple shell interactions until much higher masses (45\,M$_\odot$).  

While future work will be needed to fully assess the impact of the SH21 CBM, one expects interesting effects like unusual nucleosynthesis including more common or enhanced $i$- and $\gamma$-process nucleosynthesis. Furthermore, SN precursors and a significant change to the pre-SN structure are also expected, with many models not having the commonly expected onion-ring like structure and having a different explosion probability.

\section*{Acknowledgements}
This work is guided by the 3D hydrodynamic simulations run on the DiRAC Memory Intensive service (Cosma8) at Durham University, managed by the Institute for Computational Cosmology on behalf of the STFC DiRAC HPC Facility (www.dirac.ac.uk). The DiRAC service at Durham was funded by BEIS, UKRI and STFC capital funding, Durham University and STFC operations grants. DiRAC is part of the UKRI Digital Research Infrastructure.
EW acknowledges receipt of an STFC postgraduate studentship.
BM acknowledges support from the Australian Research Council through Discovery Project DP240101786. BM acknowledges computer time allocations from Astronomy Australia Limited's ASTAC scheme and the National Computational Merit Allocation Scheme (NCMAS). Some of this work was performed on the Gadi supercomputer with the assistance of resources and services from the National Computational Infrastructure (NCI), which is supported by the Australian Government,
RH acknowledges support from the World Premier International Research Centre Initiative (WPI Initiative), MEXT, Japan, the IReNA AccelNet Network of Networks (National Science Foundation, Grant No. OISE-1927130) and the Wolfson Foundation.
FR is a fellow of the Alexander von Humboldt Foundation. FR acknowledges support by the Klaus Tschira Foundation, by the INAF Mini grant 2024, “GALoMS - Galactic Archaeology for Low Mass Stars” (1.05.24.07.02), and by the National Recovery and Resilience Plan (NRRP), funded by the European Union - NextGenerationEU, Project ‘Cosmic POT’ (PI: L. Magrini) Grant Assignment Decree No. 2022X4TM3H by the Italian Ministry of University and Research (MUR).

\section*{Data Availability}
The key data are contained in the various tables in the paper. All the progenitor models calculated in this work are published in the Zenodo repository. Other data used or generated for this article can be shared upon reasonable request to the corresponding author.



\bibliographystyle{mnras}
\bibliography{bibliography} 




\appendix

\section{Compactness parameter taken at different mass coordinates}

\label{Appendix_compactness}
\begin{figure*}
    \includegraphics[width=\textwidth]{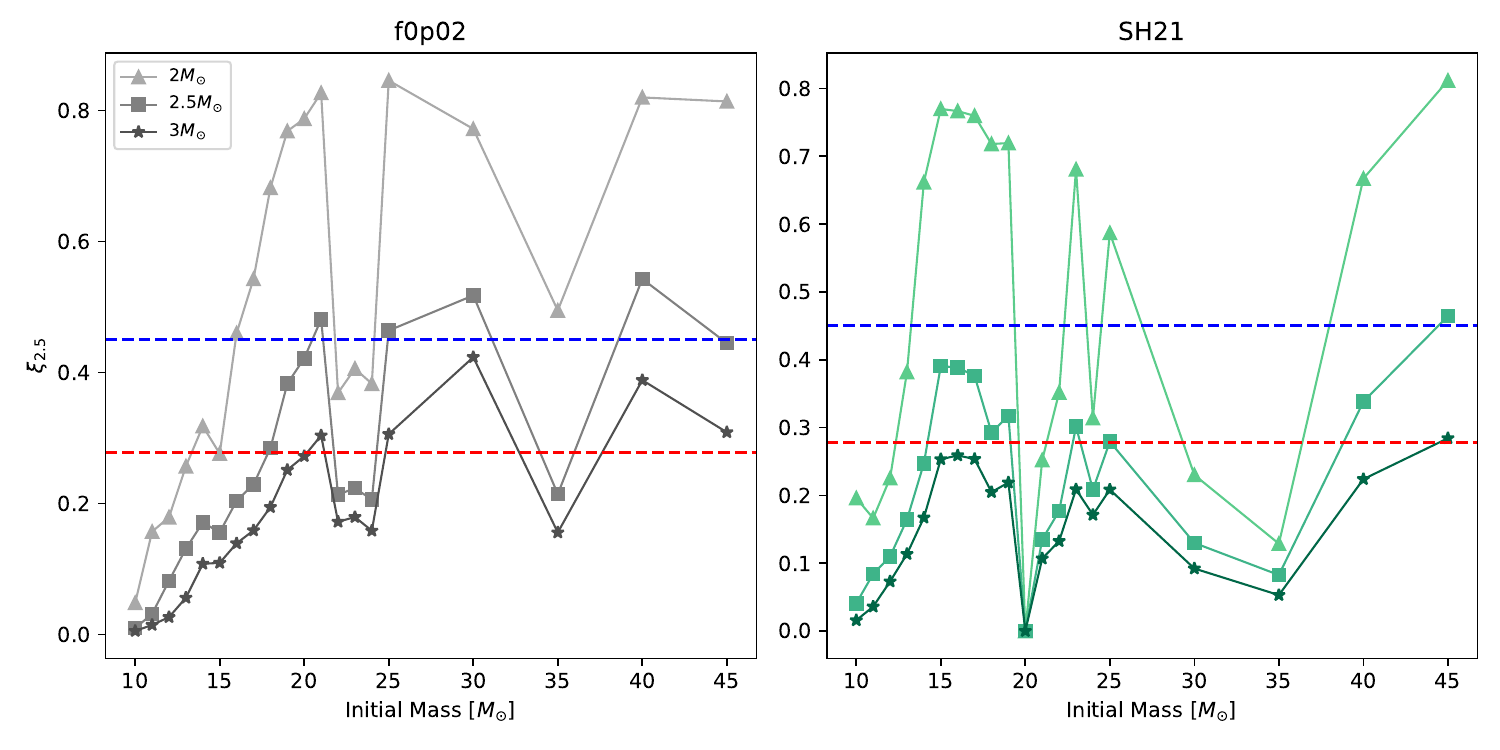}
    \caption{Comparison of compactness taken at a range of mass coordinates as a function of initial mass with the f0p02 grid shown on the left and the SH21 grid on the right. The red and blue dashed lines, similarly to Figure \ref{fig:Compactness_Subplot}, correspond to the `switch' values of compactness suggested by \citet{BM2016} and \citet{Ott2011} respectively.}
    \label{fig:masscoord}
\end{figure*}

The value of the compactness parameter changes with the mass coordinate it is measured at. In some cases, this is enough to change the predicted fate of the star, for example, the 15-19M${_\odot}$ and 40M${_\odot}$ models in the SH21 grid. When the compactness is taken at a mass coordinate of 2M${_\odot}$, both \citet{Ott2011} and \citet{BM2016} predict that the progenitor will not successfully explode. If the compactness parameter is evaluated at 2.5M${_\odot}$, then \citet{Ott2011} predicts a successful supernova whereas \citet{BM2016} predicts that it will fail. Finally, if the compactness is measured at 3M${_\odot}$, both bifurcation values predict an unsuccessful supernova. However, the general behaviour of the compactness parameter remains similar at all three mass coordinates. The reason behind measuring the compactness parameter at 2.5M${_\odot}$ is that this mass coordinate includes not only the iron core, but also the silicon and oxygen rich regions immediately surrounding it that will accrete onto the proto-neutron star during core collapse \citep{Sukhbold2014}. 


\bsp	
\label{lastpage}
\end{document}